\documentclass[runningheads]{llncs}
\usepackage[dvipsnames]{xcolor}
\usepackage{adjustbox}
\usepackage{float}
\usepackage{pgfplots}
\usepackage[backgroundcolor=Apricot]{todonotes}

\usepackage{xspace}
\usepackage{graphicx}
\usepackage[justification=centering]{caption}
\usepackage{arydshln}
\usepackage{booktabs}

\usepackage{listings}
\usepackage{amsmath}
\usepackage{microtype}
\usepackage[noadjust]{cite}
\usepackage{multirow}
\usepackage{pifont}

\usepackage[T1]{fontenc}  % this is necessary for beramo to work
\usepackage[scaled=0.74]{beramono}  % our monospace font
\newcommand{\KeY}{Ke\kern-.1emY\xspace}
\newcommand{\KeYFloat}{\KeY}

\makeatletter
\newcommand{\thickhline}{%
  \noalign {\ifnum 0=`}\fi \hrule height 1pt
  \futurelet \reserved@a \@xhline
}

\newcommand{\javamath}{\texttt{java.lang.Math}}

\usepackage[firstpage]{draftwatermark}
\SetWatermarkText{\hspace*{4.5in}\raisebox{6.3in}{\includegraphics[scale=0.1]{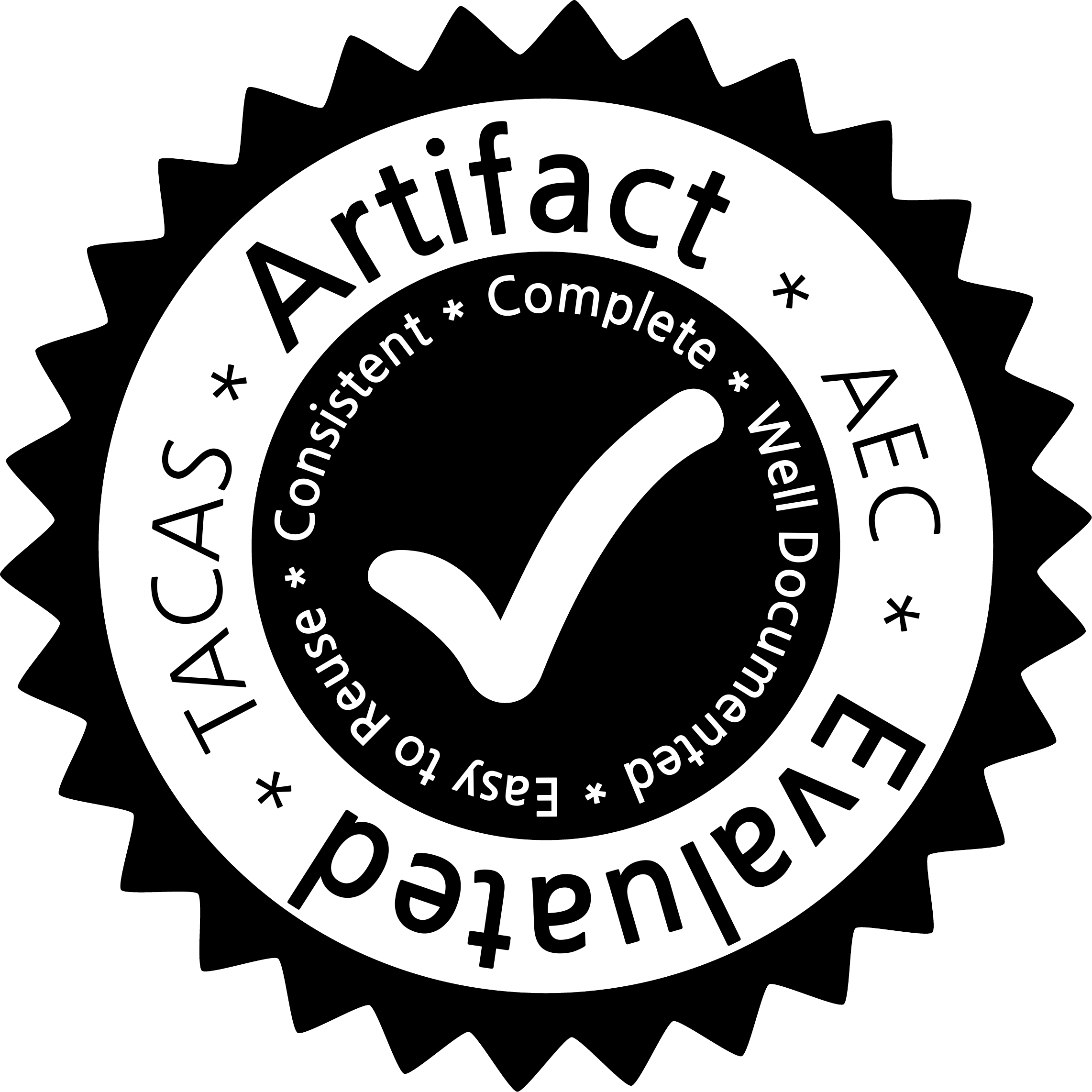}}}
\SetWatermarkAngle{0}

\makeatletter
\RequirePackage[bookmarks,unicode,colorlinks=true]{hyperref}%
   \def\@citecolor{blue}%
   \def\@urlcolor{blue}%
   \def\@linkcolor{blue}%

\def\orcidID#1{\smash{\href{http://orcid.org/#1}{\protect\raisebox{-1.25pt}{\protect\includegraphics{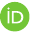}}}}}
\makeatother
\usepackage{marvosym}

\begin{document}
\title{Deductive Verification of Floating-Point\\ Java Programs in \KeY}
\author{Rosa Abbasi\inst{1}\orcidID{0000-0003-1495-3470} (\Letter)
\and Jonas Schiffl\inst{2}\orcidID{0000-0002-9882-8177}
\and Eva Darulova\inst{1}\orcidID{0000-0002-6848-3163} \and\\
  Mattias Ulbrich\inst{2}\orcidID{0000-0002-2350-1831}
  \and Wolfgang Ahrendt\inst{3}\orcidID{0000-0002-5671-2555}}
\authorrunning{R. Abbasi et al.}
\institute{MPI-SWS, Kaiserslautern and Saarbr\"{u}cken, Germany, \email{\{rosaabbasi,eva\}@mpi-sws.org} \and
Karlsruhe Institute of Technology, Karlsruhe, Germany, \email{\{jonas.schiffl,ulbrich\}@kit.edu} \and
Chalmers University of Technology, G\"{o}teborg, Sweden, \email{ahrendt@chalmers.se}
}

\maketitle              % typeset the header of the contribution
\begin{abstract}
Deductive verification has been successful in verifying interesting
  properties of real-world programs. One notable gap is the limited support
  for floating-point reasoning. This is unfortunate, as floating-point
  arithmetic is particularly unintuitive to reason about due to rounding as
  well as the presence of the special values infinity and `Not a Number' (NaN).
  In this paper, we present the first floating-point support in a deductive
  verification tool for the Java programming language. Our support in the KeY verifier handles
  arithmetic via floating-point decision procedures inside SMT solvers
  and transcendental functions via axiomatization.
  We evaluate this integration on new benchmarks, and show that this approach
  is powerful enough to prove the absence of floating-point special
  values---often a prerequisite for further reasoning about numerical
  computations---as well as certain functional properties for realistic benchmarks.

\end{abstract}

% !TEX root = main.tex
\section{Introduction}
%\todo{ @Wolfgang read over introduction}

% general intro to deductive verification
Deductive verification has been successful in providing functional
verification for programs written in popular programming languages such as
Java~\cite{KeYBook2016,MARCHE200489,DBLP:conf/nfm/JacobsSPVPP11,openjml}, Python~\cite{Nagini2018},
Rust~\cite{Prusti2019}, C~\cite{Frama-C-2012,Viper2016}, and
Ada~\cite{mccormick_chapin_2015,ChapmanSchanda2014}.
Deductive verifiers allow a user to annotate methods in a program with pre- and
postconditions, from which they automatically generate verification conditions (VCs).
These are then either proven directly by the verifier itself, or discharged with
external tools such as automated (SMT) solvers or interactive proof assistants.

While deductive verifiers fully implement many sophisticated data
representations (including heap data structures, objects, and ownership),
support for floating-point numbers remains rather limited -- solely Frama-C and
SPARK offer automated support for floating-point arithmetic in C and
Ada~\cite{Fumex2017}.
This state of affairs is at least partially a result of previous limitations
in floating-point support in SMT solvers. Consequently, deductive
verification has been used for floating-point programs only by experts with
considerable manual effort~\cite{Fumex2017,Boldo2009}.
This is unfortunate as it makes deductive verification unavailable for a
large number of programs across many domains including embedded systems,
machine learning, and scientific computing.
With the increasing need for parallelization in code, scientific
computing specifically has recently experienced algorithmic
challenges for which formal methods may contribute to a
solution~\cite{10.1145/1146238.1146256,BeckertNKSU18}.

% why floating-point reasoning is tricky
One of the main challenges of floating-point arithmetic is its
unintuitive behavior and the special values that the IEEE 754
standard~\cite{ieee75408} introduces. For instance, an overflow or a division
by zero results in the special value
(positive or negative) \emph{infinity}, and not a runtime exception. Similarly, invalid operations like
\texttt{sqrt(-1.0)} result in a \emph{Not a Number} (NaN) value. These
special values are problematic as seemingly straight-forward identities do not
hold (\texttt{x == x} or \texttt{x * 0.0 == 0.0}). In addition, every
operation on floating-point numbers potentially involves rounding, which compromises
familiar rules like associativity and distributivity. Hence, reasoning
support for writing correct floating-point programs is indispensable.

% what is the gap (what's missing today)
% Automated floating-point verification has considerably gained interest in the
% past years.
Abstract interpretation-based tools can prove the absence of runtime
errors and special values~\cite{Chen2008,Jeannet2009}, and bound roundoff errors
due to floating-point's finite
precision~\cite{Daisy2018,FPTaylor2015,Benz2012,Chiang2014,Goubault2013}.
SMT decision procedures~\cite{Brain2019} or SAT-based
model-checking~\cite{10.1145/1146238.1146256,DBLP:conf/cav/CordeiroKKST18}, on
the other hand, can prove intricate properties requiring bit-precise reasoning.
However, these techniques and tools largely support only purely floating-point
programs or program snippets, or analyze programs only up to a predefined depth
of the call stack. General reasoning about real-world object-oriented programs,
however, also requires support for features such as the (unbounded) heap,
necessitating different analyses which need to be combined with floating-point
reasoning.
%
%\todo[inline]{@Mattias: A paragraph/sentence on deductive verification vs bounded model checking}
%
% Automatic verification tools that analyze programs up to a fixed,
% predefined depth of the call stack have also included floating-point
% support. They allow the verification against a reference
% implementation or of simple
% assertions~\cite{}---but only up to the
% defined bound.

% why we want floats in *deductive verification*
Handling floating-points in a deductive verifier has unique
advantages. First, the deductive verification approach already comes with the
infrastructure for reasoning about complex control and data structures (like
exception handling and heap). Second, it allows one to flexibly combine the
verifier's symbolic execution reasoning with external decision procedures.
Third, depending on the theory support, the verifier or external solver may also
generate counterexamples of a property and thus help program
debugging -- something an abstract interpretation-based approach fundamentally
cannot provide.

% Handling floating-point arithmetic in a deductive verifier
% provides unique advantages. First, the deductive verification approach already
% comes with the infrastructure for reasoning about complex control and data structures
% (like exception handling and heap, respectively). It furthermore
% allows one to flexibly combine symbolic execution reasoning inside the deductive
% prover itself with external decision procedures.
% The main focus of deductive verification is on \emph{proving} program
% properties. However, depending on the theory support, the verifier or
% external solver may also generate counterexamples of a property and thus help
% program debugging---something an abstract interpretation-based approach
% fundamentally cannot provide.

% what this paper is about precisely
We report on adding floating-point support to the
\KeY{} deductive verifier, providing the first automated deductive
floating-point support for the Java programming language.
% why we focus on special values (mostly)
We focus mainly on proving the absence of the special values infinity and NaN.
While these are helpful in certain circumstances, for most applications they
signal an error. Hence, showing their absence is a prerequisite for further
(functional) reasoning. That said, our extension also allows one to express and discharge
arbitrary functional properties expressible in floating-point arithmetic,
including bounds on roundoff errors for certain programs, and bounds on
differences between two similar floating-point programs %, which we also demonstrate.

We exploit both \KeY{}'s symbolic execution and external SMT support. On the one
hand, we handle arithmetic operations by
relying on a combination of \KeY{}'s
symbolic execution to handle the heap and SMT based decision procedures to
handle the floating-point part of the VCs.
On the other hand, we support transcendental functions via axiomatization in the \KeY{} prover
itself.

Transcendental functions such as sine are a common feature
in numerical programs, but are not supported by floating-point
decision procedures. We explore two ways of supporting them soundly but
approximately, by encoding them as axiomatized uninterpreted function
symbols once directly in the SMT queries, and once in additional calculus rules
in \KeY{}.
Our evaluation shows that even though such reasoning is approximate, it is
nonetheless sufficient to prove the absence of special values in many
interesting programs.

% explain first main contribution (section 3)
We evaluate \KeY's floating-point support on a number of real-world
floating-point Java programs.
% For purely arithmetic
% methods, verification is fully automated, relying on a combination of \KeY{}'s
% symbolic execution to handle the heap and SMT based decision procedures to
% handle the floating-point part of the VCs.
%
Our benchmark set allows us to evaluate recent progress in SMT floating-point
support in Z3~\cite{Z32008}, CVC4~\cite{CVC42011} and
MathSAT~\cite{mathsat5} on yet unseen benchmarks. For instance, we observe
that quantifiers are challenging even if they do not affect satisfiability of
SMT queries. Our benchmarks are openly available, and we expect our
insights to be useful for further solver development.

% %
% We encode transcendental functions as uninterpreted function symbols,
% and axiomatize them using a number of relevant lemmata. Then we
% (1)~add these as axioms to the SMT queries, and, alternatively,
% (2)~implement them as additional calculus rules in \KeY{} directly.

\paragraph{Contributions}
In summary, we make the following contributions:
\begin{itemize}
	\item we implement and evaluate the first automated deductive
	verification of floating-point Java programs by
	combining the strength of rule based and SMT based deduction;
	\item we collect a new set of challenging real-world floating-point
	benchmarks in Java (available at \url{https://gitlab.mpi-sws.org/AVA/key-float-benchmarks/});
	\item we compare different SMT solvers for discharging floating-point
	VCs on this new set of benchmarks;
	\item and we develop novel automated support for reasoning about
	transcendental functions in a deductive verifier.
\end{itemize}

% !TEX root = main.tex
\section{Background}
\subsection{Introduction to \KeY}
\KeY{} \cite{KeYBook2016} is a platform for deductive verification of Java
programs, working at a source code level. The input is a Java program
annotated in the Java Modeling Language (JML) \cite{leavens2006preliminary},
encouraging a \textit{Design by Contract} (\cite{meyer1992applying,leavens_design_2006})
approach to software development.
The user specifies the expected behavior of Java classes with
\emph{class invariants} that the program has to maintain at critical points.
Methods are specified with \emph{method contracts}, consisting mainly of pre-
and postconditions, with the understanding that if the precondition holds when
the method is called, the postcondition has to hold after the method returns.

After loading an annotated program, \KeY{} translates it to a formula in Java
Dynamic Logic~\cite{KeYBook2016} (JavaDL), an instance of Dynamic
Logic~\cite{harel_dynamic_2001} which enables logical reasoning about Java
programs. Logical rules are provided for the translation of programs into
first-order logic, and for closing the resulting {\it goals}, or proof
obligations. \KeY{} is semi-interactive in that it allows manual rule
application, while also offering powerful built-in automation and macros. In
addition, it is also possible to translate an open goal into SMT-LIB format
\cite{barrett2010smt} and call an external SMT solver. For specific theories,
SMT solvers can be much more efficient than \KeY's own automation. This makes it
possible to prove some goals, which depend on SMT supported theories, by using
an SMT solver, while others are proved internally, using \KeY's own automation.

\subsection{Floating-Point Arithmetic in Java}

%Among the primitive (i.e., non-reference) types of Java, there are two which represent floating-point numbers, namely \texttt{float} and \texttt{double}. They are associated with the 32-bit and 64-bit format, respectively, as specified in the IEEE 754 Standard for Floating-Point Arithmetic.
% More precisely, Java implements a subset of that standard.
In the following, we summarize some central characteristics of Java floating-point numbers, loosely following \cite{HandbookFP}.
%Most of this is not specific to Java, but more generally applies to IEEE 754. We will often write `floating-point numbers' when we actually mean `IEEE 754 floating-point numbers' or `Java floating-point numbers'. Also note that the Java Virtual Machine (JVM) only supports floating-point numbers with base 2, even if the Java language supports base 10 as well, in such a way that parsing as well as input/output routines translate back and forth between the bases.
%
Each \emph{normal} floating-point number $x$ can be represented as a triplet
$(s,m,e)$, such that $x = (-1)^s * m * 2^e$, where $s \in \{0,1\}$ is the
\emph{sign}, $m$ (called \emph{significand}) is a binary fixed-point number with
one digit before the radix point and $p-1$ digits after the radix point (note
that $0 \leq m < 2$), and $e$ (\emph{exponent}) is an integer such that $e_{min}
\leq e \leq e_{max}$.
Java supports two floating-point formats (both in base $2$): \texttt{float}
(`single') precision with $p = 24$, and minimal and maximal exponent
$e_{min} = -126, e_{max} = 127$ and
\texttt{double} precision with $p = 53, e_{min} = -1022, e_{max} = 1023$.

Whenever the result of a computation cannot be exactly represented with the
given precision, it is rounded. IEEE 754 defines various rounding modes, of
which Java only supports \emph{round to nearest, ties to even}. Rounding is
exact, as if one would first compute the ideal real number, and round
afterwards.

The triple representation gives us two zeros, $+0$ and $-0$, represented by $(0,0,0)$ and $(1,0,0)$, respectively.
% \footnote{More precisely, the second element of these tuples has one zero before the radix point, and $p-1$ zeros after the radix point.}
If the absolute value of the ideal result of a computation is too small to be representable as a floating-point number of the given format, the resulting floating point number is $+0$ or $-0$.
In addition, there are three special values, $+\infty$, $-\infty$, and NaN (Not a Number). If the absolute value of the ideal result of a computation is too big to be representable as a floating-point number of the given format, the result is $+\infty$ or $-\infty$. Also, division by zero will give an infinite result (e.g., $7.13/+0 = +\infty$). Computing further with infinity may give an infinite result (e.g., $+\infty + +\infty = +\infty$), but may also result in the additional `error value' NaN (e.g., $+\infty - +\infty = \text{NaN}$). % Note that,
Due to the presence of infinities and NaN, floating-point operations do \emph{not} throw Java exceptions.
%\footnote{The notion of exceptions in the IEEE 754 Standard is not to be confused with Java exceptions. IEEE 754 exceptions are rather `sticky flags' for overflow, underflow, and other problematic situations. They do not change the control flow, and certainly are not objects. In this context, we can also mention that IEEE 754 exceptions are not supported by Java.}

%In Java, there are two different ways of treating floating-point computations.
By default, the Java virtual machine is allowed to make use of higher-precision formats provided by the hardware. This can make computation more accurate, but it also leads to platform dependent behaviour.
% non-determinism, because the result of the computation may depend on the hardware the program runs on.
This can be avoided by using the {\tt strictfp} modifier, ensuring that only the single and double precision types
% specified in the IEEE 754 Standard
are used. This modifier ensures portability.

% !TEX root = main.tex
\section{Floating-Point Support in \KeY}
\subsection{Arithmetics}\label{Sec:arithmetic}
In order to be able to specify and verify programs containing floating-point
numbers, we made several extensions to the \KeY{} tool.
First, we added the {\tt float}  and {\tt double} types to the \KeY type system,
together with an enum type for
the different rounding modes of the IEEE 754 Standard.

We further introduced functions and predicate symbols to formalize operations ({\tt
  +}, {\tt *}, \dots) and comparisons ({\tt <}, {\tt ==}, \dots) on
floating-point expressions. The translation supports both code with and
without the \texttt{strictfp} modifier. However,  since the actual precision of
non-strictfp operations is  not  known, the function symbols remain
uninterpreted.
We extended \KeY's parser to correctly handle programs and annotations
containing floating-point numbers, and added logic rules for translating floating-point
expressions from Java or JML to JavaDL.

\begin{lstlisting}[float, caption=The Rectangle.scale benchmark, label=rectangle-bench, linewidth=\columnwidth,breaklines=true]
/*@ public normal_behavior
  @  requires \fp_nice(arg0.x) && \fp_nice(arg0.y)
  @    && \fp_nice(arg1) && \fp_nice(arg2);
  @  ensures !\fp_nan(\result.x) && !\fp_nan(\result.y) &&
  @   !\fp_nan(\result.width) && !\fp_nan(\result.height);
  @ also
  @ public normal_behavior
  @  requires -5.53 <= arg0.x && arg0.x <= -3.38 &&
  @    -5.53 <= arg0.y && arg0.y <= -3.38 &&
  @    3.1 < arg0.width && arg0.width <= 3.7332 &&
  @    3.0000001 < arg0.height && arg0.height <=4.0004 &&
  @    3.0003001 < arg1 && arg1 <= 4.0024 &&
  @    -6.4000003 < arg2 && arg2 <= 3.0001;
  @  ensures !\fp_nan(\result.x) && !\fp_nan(\result.y)&&
  @   !\fp_nan(\result.width) &&!\fp_nan(\result.height);
  @*/
public Rectangle scale(Rectangle arg0, double arg1, double arg2){
  Area v1 = new Area(arg0);
  AffineTransform v2 = AffineTransform.getScaleInstance(arg1, arg2);
  Area v3 = v1.createTransformedArea(v2);
  Rectangle v4 = v3.getRectangle2D();
  return v4;
}
\end{lstlisting}

As an example, \autoref{rectangle-bench} shows JML specifications of our \texttt{Rectangle}
benchmark  that contains floating-point literals and makes use of the \texttt{fp\_nan} and \texttt{fp\_nice} predicates. \texttt{fp\_nan} states that a floating-point expression is NaN and
 {\tt fp\_nice}, which is shorthand for ``not infinity and not NaN'',  states that a floating-point expression is not NaN or infinity.
The \texttt{scale} method contains two contracts that are checked separately, ensuring that the class fields of a scaled rectangle object are not NaN, considering different preconditions. For the first contract, the SMT solver produces a counterexample. In the second, we bound inputs by concrete ranges that we picked arbitrarily and get the valid result. In practice, such ranges would come from the context, e.g. from the kind of rectangles that appear in an application, or from known ranges of sensor values.
%Note that \KeY checks automatically that the function is only called with valid inputs.

%As an example, \autoref{rotation-bench} shows JML specifications of our \texttt{Rotation}
%benchmark that contains floating-point literals and makes use of the \texttt{fp\_nan} and
%\texttt{fp\_infinite} predicates, which state that a floating-point expression is NaN or infinity,
%respectively. Note that in the postcondition, we use the {\tt fp\_nice} predicate, which is
%shorthand for ``not infinity and not NaN''. The postcondition also specifies that the difference
%between the original vector and the result must not be larger than a certain bound. In practice,
%such bounds would be derived from domain-specific knowledge of a programmer, or obtained from the
%results of other programs.

Concerning discharging the resulting proof obligations, there were two main
ways to consider. One is to create a floating-point theory within \KeY{} by
adding axioms and deduction rules, so that the desired properties can be
proven in \KeY's sequent calculus. The other way is to translate the proof
obligations from JavaDL to SMT-LIB and call an external SMT solver. While the
\KeY{} approach traditionally favors conducting proofs within \KeY, for this
work, we partially deviated from this way in order to harness the greater
experience and efficiency of SMT solvers when it comes to floating-point
arithmetic. Our approach attempts to get the best of both worlds by
distinguishing between basic floating-point arithmetic, i.\,e., elementary operations
and comparisons, and more complex functions which do not have an SMT-LIB
equivalent (e.\,g., the transcendental functions), or where the SMT-LIB
function is not usefully implemented by current SMT solvers (see~\autoref{sec:taclets-key}).
%(e.\,g., the square root).
% In these cases, functions become uninterpreted symbols that
% require additional axiomatization; we provide these semantic definitions
% in \KeY by adding corresponding taclets (see~\autoref{sec:taclets-key}).

Elementary operations and comparisons get translated to the corresponding
SMT-LIB functions. In SMT-LIB, all floating-point computations conform to the
IEEE 754 Standard. Therefore, only Java programs with the {\tt strictfp}
modifier can be directly translated to SMT-LIB without loss of correctness.

We developed a translation from \KeY's floating-point theory to SMT-LIB. In
order to integrate it into \KeY, we also overhauled the existing translation
from  JavaDL to SMT-LIB to create a new, more modular framework, which now
supports  all the features of the original translation, e.\,g., heaps and
integer arithmetic, but also floating-point expressions at the  same time.

Floating-point intricacies sometimes require extra caution. For
example, there are two different notions of equality for floats: bitwise equality and IEEE754
equality. Our implementation ensures these are distinguished correctly, and that the specification
language remains intuitive for a developer to use.

Using the translation to SMT-LIB, we can specify and prove two classes of
properties in \KeY: The absence of special values is specified using the {\tt
fp\_nan} and {\tt fp\_infinite} predicates (or the {\tt fp\_nice} equivalent).
Furthermore, one can specify \emph{functional} properties that are expressible
in floating-point arithmetic, e.g. one can compare the result of a computation
against the result of a different program which is known to produce a good
result or a reference value.

% Through the implementation of floating-point support for \KeY and translation to SMTLIB, two classes
% of properties can be specified, translated to logic and then proven by an SMT solver: The absence of
% special values is specified using the {\tt fp\_nan} and {\tt fp\_infinite} predicates (or the {\tt
% fp\_nice} equivalent) behaviour. Furthermore, the result of a computation can also be compared to a
% known error bound, or to the result of a different program which is known to produce a good result.

% !TEX root = main.tex
\subsection{Transcendental Functions}\label{Sec:trans}
% \todo[inline]{@Mattias?  I would have liked to draw more insights into how transcendental functions are supported within KeY. Specifically, the details of formalism are mostly excluded. As a result, though there is some intuition on how the support is built, it is not clear what is supported and what is not(from reviews)}
Floating-point decision procedures in SMT solvers successfully handle programs
consisting of arithmetic and square root operations. Many numerical real-world
programs, however, include transcendental functions such as \texttt{sin} and
\texttt{cos}. In Java programs, these functions are implemented as static library
functions in the class \javamath.

Unlike arithmetic operations, transcendental functions are much more
loosely specified by the IEEE 754 Standard---only an upper bound on the roundoff
error is given. Libraries are thus free to provide different implementations,
and even tighter error bounds. Exact reasoning in the same spirit as
floating-point arithmetic would thus have to encode a specific implementation.
Given that these implementations are highly optimized, this approach would be
arguably complex.
We observe, however, that such exact reasoning about transcendental functions is
often not necessary and a sound approximate approach is sufficient and
efficient.

In this section, we introduce an axiomatic approach for reasoning about programs
containing transcendental functions. We observe that with the flexibility of
deductive verification and \KeY itself, we can instantiate it in two different
ways. We encode transcendental functions as uninterpreted functions and
axiomatize them in the SMT queries. Alternatively, we encode these axioms in
\KeYFloat as logical inference rules.
%(so-called taclets), which can be applied selectively on the proof
%obligations before producing the SMT-LIB translation.
%
%In the following sections, we will explain each of these solutions in more
%detail and later we will evaluate them on a set of benchmarks.

\subsubsection {(A)~\textbf{Axiomatization in SMT}}
\makeatletter
\edef\@TransLabel{\relax\@currentlabel}
\edef\@currentlabel{\@TransLabel.A}
\makeatother
% better hide label since there is no number for this anyway %
\label{Sec:axiomInSolver}

%With the SMT solvers' lack of support for transcendental functions, we designed a
%sound but approximate solution to be able to reason about Java programs
%containing calls to library functions.
%
We encode library functions as uninterpreted functions and include a set of axioms
in the SMT-LIB translation for each method that is called in a benchmark.
That is, we extended \KeY such that when a transcendental function exists in the
proof obligation, its definition alongside all the axioms for that function are
added to the translation.
% We are able to do this since we have already encoded such methods as uninterpreted
% functions in \KeY and thus  \KeY is able to apply Taclet rules on terms that
% include such methods and successfully create the proof obligations.

%The axioms are expressed as quantified floating-point formulas and capture
%high-level properties of library functions complying with the specifications in
%the IEEE 754 Standard.
For the axiomatization of transcendentals, we did \emph{not} add rules
that expand to a definition or allow a repeated approximation of the
function value (like expansion into a Taylor series). Instead, we
added a number of lemmata encoding interesting properties related to special values.
%relevant to the solution of the verification conditions that we aimed at.
%In particular, they encode value ranges and allow, in particular,
%one to show that a function application is not NaN.
For instance, the following axiom states that if the input to the
\texttt{sin} function is not a NaN or infinity, then the returned value of
\texttt{sin} is between $-1.0$ and $1.0$:
\begin{lstlisting} %[mathescape=true,caption=Example axiom specifying $sin(a) \in [-1, 1]$,label=axiomExample]

(assert (forall ((a Float64)) (=>
  (and (not (fp.isNaN a)) (not (fp.isInfinite a)))
  (and (fp.leq (sinDouble a) (fp #b0 #b01111111111 #b0000...000000))
       (fp.geq (sinDouble a) (fp #b1 #b01111111111 #b0000...000000))))))
\end{lstlisting}
% Listing~\ref{axiomExample} shows one such axiom for the \texttt{sin}
% function in the SMT-LIB format. This axiom states that if the input argument
% of the \texttt{sin} function is not a NaN or infinity, then the returned value of
% \texttt{sin} is between $-1.0$ and $1.0$.
Note that this implies that the result is not a NaN or infinity.
The other axioms are similar in spirit, so we do not list them.

These axioms are expressed as quantified floating-point formulas and capture high-level properties of library functions complying with the specifications in
the IEEE 754 Standard.
Clearly, since we do not have the actual implementations of these functions, we
are not able to prove arbitrary properties. However, such an axiomatization is
often sufficient to check for the (absence of) special values, i.e. NaN and
infinity, as our experiments in~\autoref{Sec:trans-eval} show.

%\begin{adjustbox}{width=1\textwidth}
%\end{adjustbox}

\subsubsection{\textbf{(B)~Taclets in \KeY}}
\makeatletter
\edef\@currentlabel{\@TransLabel.B}
\makeatother
\label{sec:taclets-key}

% \todo{@Mattias} \mattias{make sure/check that the first paragraph can
% 	go with the last subsections! Perhaps some of it goes to discussion}
Reasoning about quantified formulas in SMT is a long-lasting challenge \cite{GeM09ext}.
We have also observed in our experiments with only arithmetic operations
(\autoref{Sec:arith-eval}) that SMT solvers struggle with quantifiers in
combination with floating-points.
We have therefore implemented an alternative approach encoding the axioms not in the
SMT queries, but instead as deductive inference rules (so-called taclets) in \KeY.

The rules encode the same logical information as the
universally quantified assertions that we add in SMT-LIB (and where we leave the
choice of instantiations entirely to the SMT/SAT solver).
With our taclet approach, we instantiate a quantifier (only) to one's needs. We
note that for proving a property correct, this results in a correct
(under)approximation. However, the prize for achieving more closed proofs and shorter running times is
that for disproving a property, not considering all possible quantifier
instantiations may lead to spurious counterexamples, i.e., false positives.

A heuristic strategy applies the rules automatically using the
occurrences of transcendentals as instantiation triggers. However,
instantiating the axioms too eagerly, considerably increases the number
of open goals, which is why we assume that the user selects the axioms
to apply manually (and did so in the experiments).  After the
application the proof obligation can either be closed, i.e proven, by \KeY{}
automatically, or be given to the SMT solver as before for final
solving.

Currently, the set of axioms (in the SMT-LIB translation and
as taclets in \KeY) only contains axioms for the
transcendental functions occurring in our benchmarks. So far we have $10$ axioms;
however, adding more axioms (also for further transcendentals like
exponentiation or logarithm) is straightforward. The full set of axioms is included in the Appendix of the technical report.

% \todo[inline]{@Mattias, Please explain what is involved in automating this step. (from the reviews)}

% \begin{displaymath}
% \begin{array}{l@{~~}l}
% \mathtt{find}&\mathit{sin(a)} \\
% \mathtt{add}&\mathit{\lnot fp\_nan(a) \land
% 	\lnot fp\_infinite(a)\to} \\
% \mathtt{}&\mathit{-1.0 \leq sin(a) \leq +1.0
% 	~~\Longrightarrow}
% \end{array}
% \end{displaymath}
%
%\tacletA{\mathit{sin(a)}} {\mathit{\lnot fp\_nan(a) \land
%		\lnot fp\_infinite(a)\to -1.0 \leq sin(a) \leq +1.0
%		~~\Longrightarrow}}
%
% For instance, the taclet that corresponds to the SMT axiom listed in
% Listing~\ref{axiomExample} allows one to add the assumption that if $a$ is not NaN and
% not $\pm\infty$, then $\mathit{sin(a)}$ is between $-1.0$ and $+1.0$.
% The rule is schematic with $a$ being the parameter that is looked
% for within the sequent. Hence, knowledge about uninterpreted functions
% is added to the sequent on demand and only for selected instances; in this case, for
% any (ground) term $\mathit{sin}(t)$ occurring on the sequent.
% A similar taclet
% has been added for $-1.0 \leq sin(a) \leq +1.0$ instead of
% $\mathit{\lnot fp\_nan(sin(a))}$ to argue about the value range and
% not only about the absence of NaN.

%%% Local Variables:
%%% mode: latex
%%% TeX-master: "main"
%%% End:

% !TEX root = main.tex
\section{Evaluation}
	\subsection{Benchmark Programs}

	%%new
	\begin{table*}[t]
	\centering
	\small
	\begin{adjustbox}{width=1\textwidth}
	\renewcommand{\arraystretch}{1.1}
\begin{tabular}{lcccl|cccr}
	\toprule
	% \multirow{2}{*}{benchmark} & \multirow{2}{*}{\# classes} & \multirow{2}{*}{\# method calls} & \multirow{2}{*}{\# arithmetic ops} & \multirow{2}{*}{library functions} & \multicolumn{2}{c}{\# goals}  & \multirow{2}{*}{\begin{tabular}[c]{@{}c@{}}\# rules \\ applied\end{tabular}} & \multicolumn{1}{l}{\multirow{2}{*}{\begin{tabular}[c]{@{}l@{}}automode \\ time(s)\end{tabular}}} \\

	& \multicolumn{4}{c|}{Benchmark Details} & \multicolumn{4}{c}{Automode Statistics} \\
	benchmark &
	\# classes &
	\begin{tabular}[c]{@{}c@{}}\# method \\ calls\end{tabular} &
	\begin{tabular}[c]{@{}c@{}}\# arith.\\ ops \end{tabular}&
	\begin{tabular}[c]{@{}c@{}}library\\ functions \end{tabular}&
	\begin{tabular}[c]{@{}c@{}}\# goals closed \\by \KeY \end{tabular}&
	\begin{tabular}[c]{@{}c@{}}\# goals to be\\ closed externally \end{tabular}&
	\begin{tabular}[c]{@{}c@{}}\# rules\\ applied \end{tabular}&
	\begin{tabular}[c]{@{}c@{}}automode\\ time (s)\end{tabular} \\

	%&   &  &    &    & closed by KeY & to be closed externally &    & \multicolumn{1}{l}{}   \\ \hline
	\midrule
	Complex.add (2)& 1 & 0& 2  & -  & 3 / 3   & 1 / 4   & 185 / 286 & 0.7 / 0.2   \\
	Complex.divide (2)  & 1 & 0& 11 & -  & 10 / 8  & 2 / 8    & 483 / 625 & 0.7 / 0.8\\
	Complex.compare  & 1 & 0& 2  & -  & 3   & 2    & 216 & 0.2 \\
	Complex.reciprocal (2)& 1 & 1& 6  & -  & 1 / 1   & 2 / 2    & 402 / 406 & 0.4 / 0.5\\
	Circuit.impedance& 2 & 1& 3  & -  & 1   & 4    & 360 & 0.5    \\
	Circuit.current (2)  & 2 & 3& 14 & -  & 11 / 11  & 4 / 1    & 1267 / 1238& 4.0 / 4.1  \\
	Matrix2.transposedEq       & 1 & 3& 3  & - & 3   & 1    & 735 & 0.9 \\
	Matrix3.transposedEq       & 1 & 4& 34 & -  & 3   & 1    & 1786& 5.1 \\
	Matrix3.transposedEqV2     & 1 & 4& 34 & -  & 3   & 1    & 1796& 5.4 \\
	Rectangle.scale (2)& 3 + 1       & 23     & 22 & -  & 32 / 32  & 32 / 16   & 5990 / 5617& 18.4 / 14.5    \\
	Rotate.computeError& 1 + 1      &  6    & 26 &  -  & 108  & 8   & 3693 &   74.2    \\
	Rotate.computeRelErr& 1 + 1     & 6     & 28 &  - &120   & 8   & 3898 & 79.6     \\
	FPLoop.fploop          & 1          & 0     & 1     & -    & 2    & 4   & 99     & 0.1     \\
	FPLoop.fploop2         & 1          & 0   & 1     & -     & 2    & 4   & 99     & 0.1     \\
	FPLoop.fploop3         & 1          & 0   & 1   & -      & 2    & 4    & 99     & 0.1      \\
	Cartesian.toPolar& 2 + 1       & 3& 6  & sqrt, atan& 1   & 4       & 438 & 0.5     \\
	Cartesian.distanceTo       & 1 + 1       & 1& 5  & sqrt& 2   & 1       & 191 & 0.1 \\
	Polar.toCartesian& 2 + 1       & 3& 4  & cos, sin& 1   & 2       & 364 & 0.5  \\
	Circuit.instantCurrent     & 2 + 1       & 14     & 23 & sqrt, atan, cos    & 17  & 2       & 1686& 14.1 \\
	Circuit.instantVoltage     & 1 + 1       & 1& 4  & cos     & 0   & 2       & 138 & 0.1\\
	\bottomrule
\end{tabular}
	\end{adjustbox}
	%\captionsetup{width=.9\textwidth}
	\caption{Benchmark details and \KeYFloat automode statistics, time is measured in seconds}
	\label{Tab:benchmarkStatistics}
\end{table*}

We collected a set of existing floating-point Java programs representing
real-world applications in order to evaluate the feasibility and performance of
\KeY's floating-point support.

The left half of \autoref{Tab:benchmarkStatistics} provides an overview of our benchmarks.
Each benchmark consists of one method, which is composed of arithmetic
operations and method calls to potentially other classes.
The invocations of methods from \javamath{} (e.g. \texttt{Math.abs}) are
marked by ``+1'' in~\autoref{Tab:benchmarkStatistics}; these are resolved by
inlining the method implementation. For benchmarks that contain calls to
transcendental functions and square root, the called functions are listed; these
are handled by our axiomatization.
We include \texttt{sqrt} in this list, as we have observed that exact support can be expensive, so
it may be advantageous to handle \texttt{sqrt} axiomatically. Benchmarks  \texttt{Rectangle}, \texttt{Circuit}, \texttt{Matrix3} and \texttt{Rotation} are partially shown in
Listings \ref{rectangle-bench}, \ref{circuit-bench}, \ref{matrix-bench} and \ref{rotation-bench}  respectively.

Each benchmark also includes a JML contract that is to be checked. For some methods, we specify two contracts (marked by ``(2)'' in the first column of \autoref{Tab:benchmarkStatistics}), each serving as an independent benchmark.
The contracts for most of these benchmarks check that the methods do not return a
special value i.e infinity and/or NaN, the preconditions being that the
variables are not themselves special values and possibly are bounded in a given
range.
For the \texttt{Matrix}, \texttt{FPLoop} and \texttt{Rotate} benchmarks, we check a
\emph{functional} property (see~\autoref{sec:func-prop}).
\texttt{FPLoop}, which has three contracts, additionally shows how to specify floating-point loop
behavior using loop invariants.

\begin{lstlisting}[float, caption=The Circuit.instantCurrent benchmark, label=circuit-bench]
public class Circuit {
double maxVoltage, frequency, resistance, inductance;
// ...

/*@ public normal_behavior
  @ requires 1.0 < this.maxVoltage && this.maxVoltage < 12.0 &&
  @  1.0 < this.frequency && this.frequency < 100.0 &&
  @  1.0 < this.resistance && this.resistance < 50.0 &&
  @  0.001 < this.inductance && this.inductance < 0.004 &&
  @  0.0 < time && time < 300.0;
  @ ensures !\fp_nan(\result) && !\fp_infinite(\result);
  @*/
public double instantCurrent(double time) {
  Complex current = computeCurrent();
  double maxCurrent = Math.sqrt(current.getRealPart() * current.getRealPart() +
    current.getImaginaryPart() * current.getImaginaryPart());
  double theta = Math.atan(current.getImaginaryPart() / current.getRealPart());
  return maxCurrent * Math.cos((2.0 * Math.PI * frequency * time) + theta);
}}
\end{lstlisting}
\subsection{Proof Obligation Generation}

To reason about the contract of a selected benchmark, we apply \KeYFloat,
which generates proof obligations or `goals'. Some of these goals (heap-related) are closed  by \KeYFloat automatically. The remaining open goals are closed by either SMT solvers with floating-point support directly
(\autoref{Sec:arithmetic} and \autoref{Sec:axiomInSolver}), or with a combination of transcendental \KeY
taclets and floating-point SMT solving (\autoref{sec:taclets-key}).

Columns 6 and 7 in \autoref{Tab:benchmarkStatistics} show the number of proof obligations closed by \KeYFloat directly and
to be discharged by external solvers, respectively.
The next two columns show the number of taclet rules that \KeYFloat applied in
order to close its goals, and the time this takes.
For benchmarks with two contracts we show the respective values separated by `/'.

We run our experiments
on a server with 1.5\,TB memory and 4x12 CPU cores at 3\,GHz.
However, \KeY runs single-threadedly and does not use more than 8GB of memory.

For our set of benchmarks, the symbolic execution process is fully automated.
Note that the machinery can deal with loop invariants, if they are provided.
%We have a benchmark that includes loops which requires and contains loop invariants.
Loop invariant generation is, however, particularly challenging for floating-points
due to roundoff errors~\cite{Darulova2017,Izycheva20}, and a research topic in itself.
%%TODO add reference to SAS20

%%1-6 fixed
\subsection {Evaluation of SMT Floating-Point Support}\label{Sec:arith-eval}

%While \KeYFloat fully automatically closes goals related to the heap
%structure, the reasoning about floating-point arithmetic itself is discharged
%externally by SMT solvers with dedicated decision procedures.
Previous work~\cite{Fumex2017} reported that SMT support for floating-point
arithmetic is rather limited. However, with recent advances~\cite{Brain2019},
we evaluate the situation again.
Most benchmarks used to evaluate SMT solvers' decision
procedures~\cite{QFBench} aim to check (individual) specialized (corner case)
properties of floating-point arithmetic. The proof obligations generated from
our set of benchmarks are complementary in that they are more arithmetic
heavy, while nonetheless relying on accurate reasoning about special values and functional properties.

For each open goal not automatically closed, \KeYFloat generates one SMT-LIB
file that is fed to the solvers for validation. We
compare the performance of the three major SMT solvers with
floating-point support CVC4~\cite{CVC42011} (version 1.8, with the SymFPU library~\cite{Brain2019} enabled), Z3 (4.8.9)~\cite{Z32008} and MathSAT (5.6.3)~\cite{mathsat5}.
% In particular, we evaluate the solvers' support for quantified formulas, since by
%default these are part of \KeY's modular translation. We furthermore show that
%the solvers' performance can vary substantially depending on the version or
%set of parameters, as well as relatively minor changes to preconditions. We
%think that our observations can be helpful both for further decision procedure
%development and program verification.
%In the following, we examine each aspect in detail and
For this we set a timeout of 300s for each proof obligation. While \KeYFloat is able
to discharge proof obligations in parallel, for our experiments, we do so
sequentially to maintain comparability.

\KeYFloat's default translation to SMT includes quantifiers.
These quantifications are not related to floating-point
arithmetic, but are used to logically encode important properties
of the Java memory model, like the type hierarchy and the absence
of dangling references on any valid Java heap. If we reason about
floating-point problems in isolation, they are not needed,
but if we want to consider Java verification more holistically
with questions combining aspects of heap and floating point
reasoning, they become essential. We manually inspected that the proof
obligations without our axiomatized treatment of transcendental functions do not depend on these properties and investigate the quantifier support by including or removing them
from the SMT translations. We do not report results with quantifiers for MathSAT,
since it does not support them.

\begin{table*}[t]
  \centering
  \begin{adjustbox}{width=1\textwidth}
  \renewcommand{\arraystretch}{1.1}
\begin{tabular}{clcccrcrcr}
  \toprule
  \multirow{2}{*}{index} & \multirow{2}{*}{experiment}                                                     & \multirow{2}{*}{\begin{tabular}[c]{@{}c@{}}quantified\\ axioms\end{tabular}} & \multirow{2}{*}{\# goals} & \multicolumn{2}{c}{CVC4} & \multicolumn{2}{c}{Z3}   & \multicolumn{2}{c}{MathSAT} \\ \cline{5-10}
  &                                                                                 &                                                                              &                           & \# goals decided  & avg. & \# goals decided & avg.  & \# goals decided   & avg.   \\ \midrule
  1                      & \multirow{2}{*}{\begin{tabular}[c]{@{}l@{}}valid \\ contracts\end{tabular}}  & \ding{51}                                                   & 80                        & 79                & 4.1  & 25               & 18.4   & -                  & -      \\
  2                      &                                                                                 & \ding{55}                                                    & 80                        & 79                & 4.0  & 52               & 35.0   & 80                 & 8.8    \\ \midrule
  3                      & \multirow{2}{*}{\begin{tabular}[c]{@{}l@{}}invalid\\  contracts\end{tabular}} &\ding{51}                                                   & 9                         & 0                 & 3.4  & 0                & 3.4   & -                  & -      \\
  4                      &                                                                                 & \ding{55}                                                     & 9                         & 8                 & 36.7 & 7                & 27.6  & 9                  & 3.9   \\ \midrule
  5                      & axioms in SMT                                                                   & \ding{51}                                                  & 10                        & 9                 & 33.2 & 4                & 63.4     & -                  & -      \\
  6                      & axioms as taclets                                                               & \ding{55}                                                     & 10                        & 10                & 33.4 & 5                & 74.2 & 8                  & 0.9    \\ \midrule
  7                      & fp.sqrt                                                                         & \ding{55}                                                     & 7                         & 7                 & 46.2 & 1                & 23.5 & 5                  & 0.4    \\
  8                      & axiomatized sqrt                                                                & \ding{55}                                                   & 7                         & 5                 & 2.4 & 5                & 282.8  & 5                 & 5.7   \\ \bottomrule
\end{tabular}
\end{adjustbox}
\caption{Summary of  valid / invalid goals correctly decided and average running times of each solver for the SMT translations with and without quantified axioms}
\label{Tab:smtResultsWithWithoutAxioms}
\vspace{-0.5cm}
\end{table*}
\begin{figure*}[t]
  \centering
  \begin{minipage}{0.495\linewidth}
    \includegraphics[width=\linewidth]{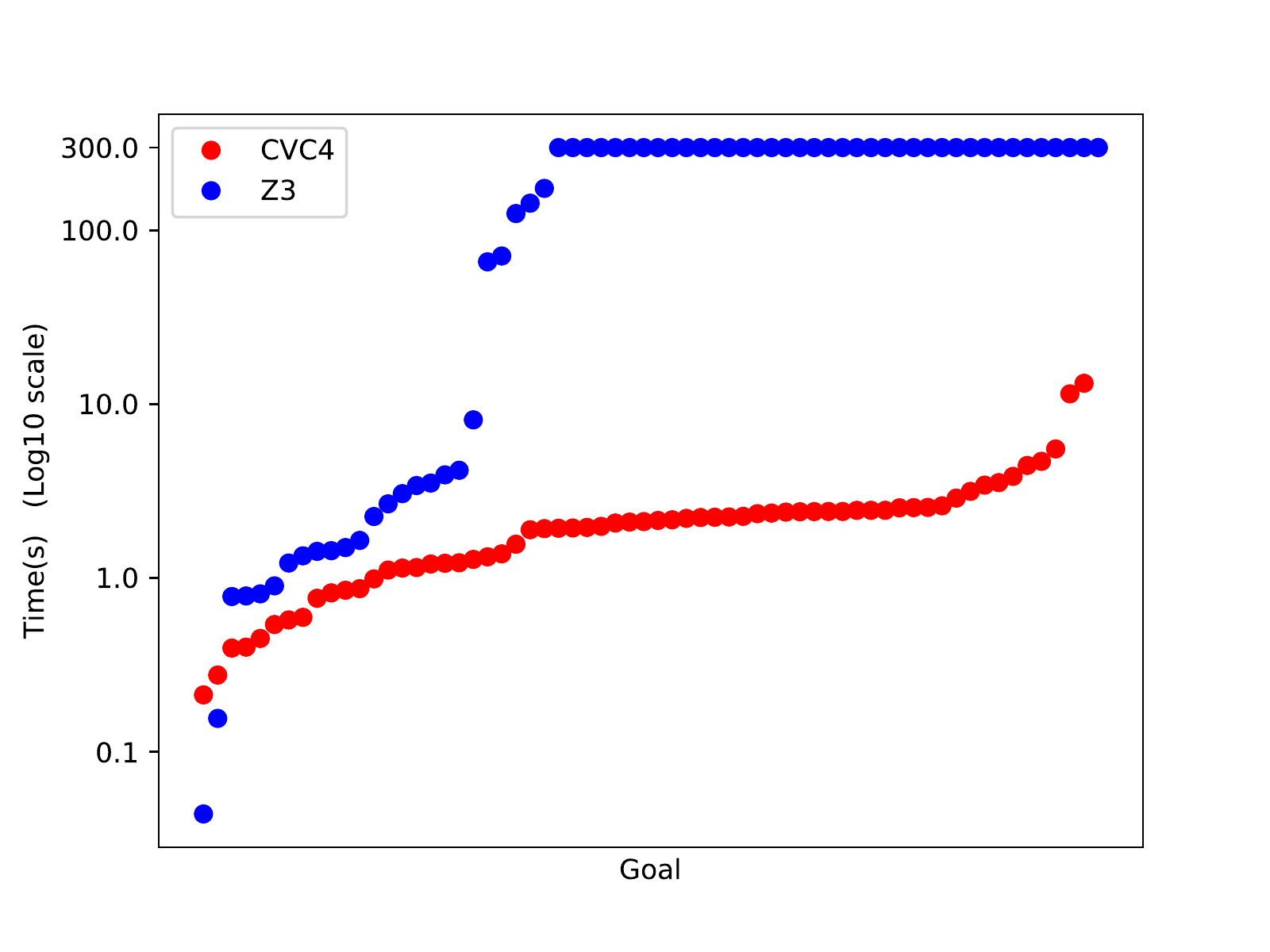}
    \caption{Runtimes for valid goals with SMT translations \emph{with} quantifiers}
    \label{Fig:vanillaWithQuantifers}
  \end{minipage}\hfill
  \begin{minipage}{0.495\linewidth}
    \includegraphics[width=\linewidth]{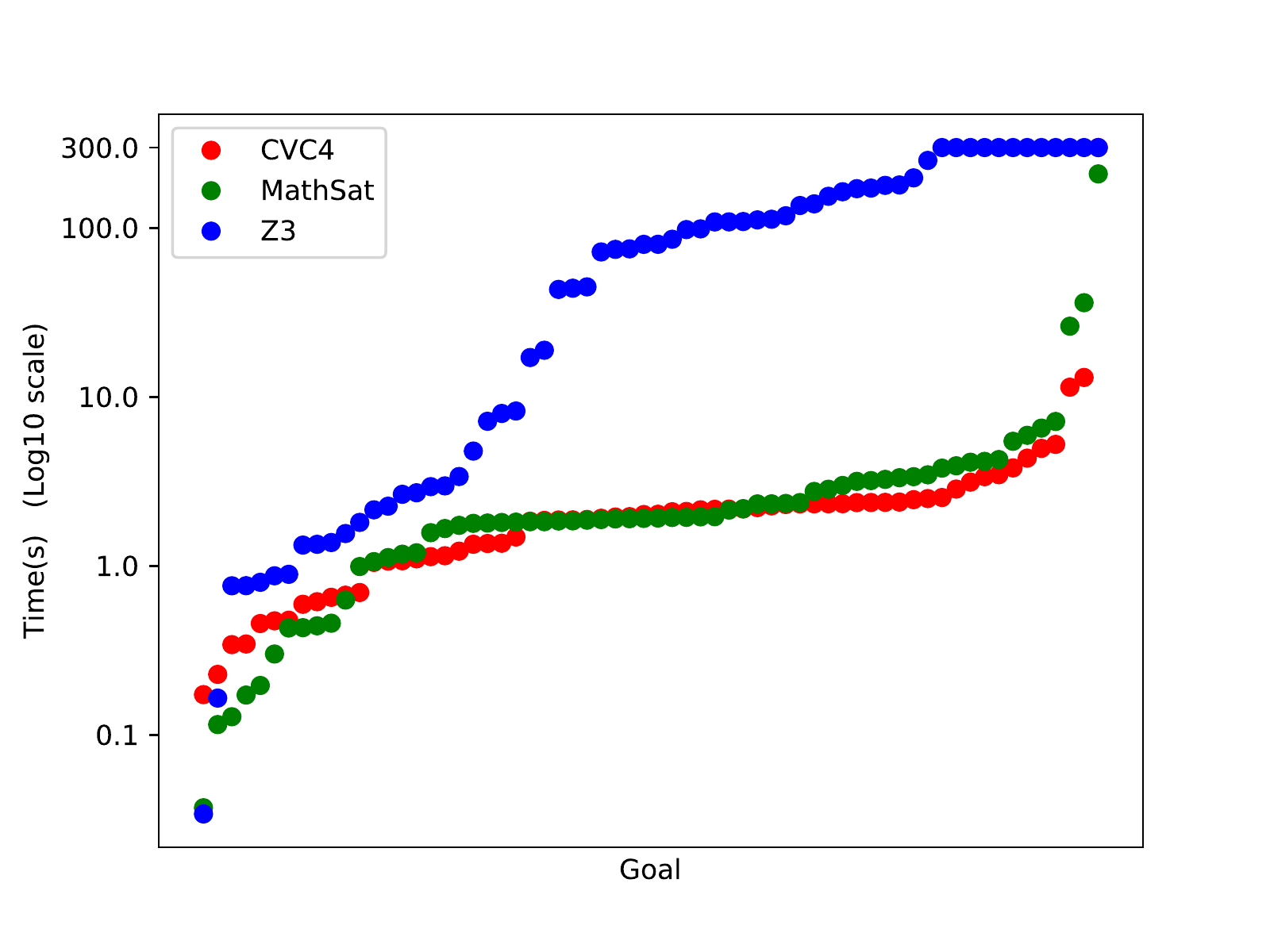}
    \captionof{figure}{Runtimes for valid goals with SMT translations \emph{without} quantifiers}
    \label{Fig:vanillaWithoutQuantifers}
  \end{minipage}
\end{figure*}

\autoref{Tab:smtResultsWithWithoutAxioms} summarizes the results of our
experiments. Column 4 shows the number of expected valid or invalid goals for
all benchmarks. For each solver we show the number of goals that each solver can
validate or invalidate, together with the average time (in seconds) needed. The
goals resulting in timeout were excluded from the computation of the average
time. Column 3 shows whether the SMT queries include quantifiers or not.

Rows 1 and 2  of \autoref{Tab:smtResultsWithWithoutAxioms} show the results for
benchmarks with valid contracts. This experiment thus represents the common
behavior of \KeYFloat, whose main goal is to \emph{prove} contracts correct.
Rows 3 and 4 of \autoref{Tab:smtResultsWithWithoutAxioms}  demonstrate the
results for benchmarks with invalid contracts, i.e. for those we expect a
counterexample for at least one of the goals.
%%Tables~\ref{Tab:smtResultsWithAxioms}, \ref{Tab:smtResultsWithoutAxioms} and
%%\ref{Tab:smtResultsCEWithWithoutAxiomsDetailed} in
The Appendix (\autoref{Sec:appendix})
%appendix of the technical report
%%TODO create the technical report
contains the
detailed results for each experiment separated by benchmark.
\autoref{Fig:vanillaWithQuantifers} and \autoref{Fig:vanillaWithoutQuantifers} show a more detailed view of the solvers' running time
for the valid benchmarks.
The x-axis shows the number of open goals that are discharged by the SMT
solvers, sorted by running time for each solver individually. The $k$-th point
of one graph shows the minimum running time needed by the solver to close each
of the $k$ fastest goals. Note that each solver may have different goals which
are its $k$ fastest. The y-axis shows the time on a logarithmic scale.

We conclude that in the presence of quantified axioms and floating-point
arithmetic solvers' performance deteriorate for both valid and invalid goals. In particular,  none of the solvers is able to find counterexamples for any of
the invalid goals. However, when the quantified axioms are
removed from the SMT translations, their performance improves.
For valid contracts, CVC4 and MathSAT perform better than Z3,  in terms of both number of goals validated and the  running time per goal.  In particular, MathSAT is able to
prove all goals. However, the running time performance of CVC4 is better than MathSAT's.
For invalid contracts, solvers are able to produce the expected counterexamples at least partially. Particularly, MathSAT has a better performance than CVC4 and Z3 in terms of both running time and the number of proof obligations for which it can produce
counterexamples.
%Another interesting observation for benchmarks with valid contracts is that the older
%version of Z3 is significantly faster than the newest version on some
%benchmarks (See  \autoref{Tab:smtResultsWithAxioms} in the Appendix).

We conducted another experiment on our \texttt{Rectangle.scale} benchmark
to assess the solvers' sensitivity to various changes,
applied to the benchmark's contract or its implementation. We considered modifications such as reducing the number of classes while keeping the same functionality, having tighter and larger bounds for variables, reducing the number of arithmetic operations etc. The details of this experiment can be found in
%\autoref{Sec:variations} in the Appendix.
the Appendix of the technical report.
In summary, solvers' performance seems to be sensitive to slight innocuous
looking changes such as the number of classes involved and variable bounds.
For example, constraining \texttt{arg2} in the original benchmark more
tightly allows CVC4 to validate all goals (1 more).
%and Z3 2 more than for the original
%benchmark.
This behavior could be potentially exploited by e.g. relaxing a variable's
bounds.

\begin{lstlisting}[float=p, caption=The Matrix3 benchmark, label=matrix-bench]
public class Matrix3 {
  double a, b, c, d, e, f, g, h, i; //The matrix: [[a b c],[d e f],[g h i]]
  double det;
  // method transpose not shown

  double determinant() {
    return (a * e * i + b * f * g + c * d * h) -
      (c * e * g + b * d * i + a * f * h);
  }
  double determinantNew() {
    return (a * (e * i) + (g * (b * f) + c * (d * h))) -
      (e * (c * g) + (i * (b * d) + a * (f * h)));
  }
  /*@ ensures \fp_normal(\result) ==> (\result == det); @*/
  double transposedEq() {
    det = determinant();
    return transpose().determinant();
  }
  /*@ ensures \fp_normal(\result) ==> (\result == det); @*/
  double transposedEqV2() {
    det = determinantNew();
    return transpose().determinantNew();
  }
}
\end{lstlisting}
\begin{lstlisting}[float=p, caption=The Rotation benchmark, label=rotation-bench, linewidth=\columnwidth,breaklines=true]
public class Rotation {
  final static double cos90 = 6.123233995736766E-17;
  final static double sin90 = 1.0;

  // rotates a 2D vector by 90 degrees
  public static double[] rotate(double[] vec) {
    double x = vec[0] * cos90 - vec[1] * sin90;
    double y = vec[0] * sin90 + vec[1] * cos90;
    return new double[]{x, y};
  }
  /*@  requires (\forall int i; 0 <= i && i < vec.length;
    @   \fp_nice(vec[i]) && vec[i] > 1.0 && vec[i] < 2.0) && vec.length == 2;
    @  ensures  \result[0] < 1.0E-15 && \result[1] < 1.0E-15;
    */
  public static double[] computeError(double[] vec) {
    double[] temp = rotate(rotate(rotate(rotate(vec))));
    return new double[]{Math.abs(temp[0] - vec[0]), Math.abs(temp[1] - vec[1])};
  }
}
\end{lstlisting}

\paragraph {Proving Functional Properties}\label{sec:func-prop}
Listings~\ref{matrix-bench} and~\ref{rotation-bench} show examples of functional
properties that are expressible in floating-point arithmetic and that \KeY can
handle. The verification results are included in rows 1 and 2 of~\autoref{Tab:smtResultsWithWithoutAxioms},
for more details see the Appendix of the technical report.
% Tables~\ref{Tab:smtResultsWithAxioms},~\ref{Tab:smtResultsWithoutAxioms}
%and~\ref{Tab:smtResultsCEWithWithoutAxiomsDetailed}.)

For \texttt{Matrix},  we check that
the determinants of a matrix and its transpose are equal. Note
 that this property holds trivially under real arithmetic, but not necessarily under
 floating-points. After feeding  \texttt{transposedEq} (which uses the \texttt{determinant} method) and its contract to \KeY, increasing the default timeout sufficiently and discharging the created goal, CVC4 generates a counterexample in 170.2s seconds and MathSAT in 16.2s. Z3 times out after 30 minutes. By feeding \texttt{transposedEqV2} (which uses the \texttt{determinantNew} method) to \KeY,  CVC4 validates the contract in 1.1s, MathSAT in 3.9s and Z3 times out again.
 One thing worth noting is that the way programs are written can greatly influence the computational complexity needed to reject or verify the contract. This is evident from the fact that slightly modifying the order of operations (using \texttt{determinantNew} instead) substantially reduces verification time and changes the verification result for MathSAT and CVC4.
 % Another thought is that the choice of SMT solver may indeed influence the performance of verification.  It seems that MathSAT and CVC4 handle some commutative expressions better than Z3 does.

 For \texttt{Rotate}, we check that the difference between an original vector and the one that is rotated four times by 90 degrees, must not be larger than 1.0E-15. We also verified the same bound for the relative difference (by exploiting another method and contract) for this benchmark. The constant \texttt{cos90} in \autoref{rotation-bench} is not precisely 0.0 to account for rounding effects in the computation of the cosine.
\texttt{FPLoop} includes three loops, for which the contracts check that the return value
is bigger than a given constant.
% We also included a benchmark with three contracts, containing loops (\texttt{FPLoop}), showcasing how to specify loop behavior in conjunction with floating-point functionality by using loop invariants. The method contracts state that the returned value of the methods should be at least as large as the threshold stated in the loops.

 Though not always very fast, these examples show that verification of
 functional floating-point properties is viable.
 % The summarized results in rows 1 and 2 of \autoref{Tab:smtResultsWithWithoutAxioms} also include the data of these two benchmarks and Tables~\ref{Tab:smtResultsWithAxioms},  \ref{Tab:smtResultsWithoutAxioms} and \ref{Tab:smtResultsCEWithWithoutAxiomsDetailed} in the Appendix, show the results for each benchmark separately.

\subsection {Evaluation of Support for Transcendental Functions in \KeY}\label{Sec:trans-eval}
We evaluated the two approaches from Section~\ref{Sec:axiomInSolver} on our set of benchmarks; rows 5 and 6 in
\autoref{Tab:smtResultsWithWithoutAxioms} summarize the results. (The detailed
results of these experiments are included in the Appendix of the technical report.)
%\autoref{Tab:axiomInKeyWithWithoutQuantifierDetailed}.)
%For both solutions, we first
%apply \KeY on a benchmark and create the proof obligations.
%Our first purely SMT-based approach is fully automated.
Note that both approaches are fully automated.

We conclude that the SMT solvers perform better when the
axiomatization is applied at the \KeY level. When axioms for transcendental functions are added to the SMT-LIB
translation directly Z3 validates 4 out of 10  goals. With the
axiomatization at the \KeY level, solvers are able to validate more goals (with
quantified formulas removed from the SMT translations), e.g. Z3 is able to
validate 5 goals and CVC4 can validate all.
%Furthermore, the average running time for CVC4 has decreased.
Therefore, it is preferable to apply them on the \KeY side via taclet rules.

%\subsubsection*{Axiomatized \texttt{sqrt} vs \texttt{fp.sqrt}}

%\begin{table*}[t]
%	\centering
%%	\begin{adjustbox}{width=1\textwidth}
%		\renewcommand{\arraystretch}{1.2}
%		\begin{tabular}{llccrcrcr}
%			\hline
%			\multirow{2}{*}{quantified axioms} & \multirow{2}{*}{approach} & \multirow{2}{*}{\# goals} & \multicolumn{2}{c}{CVC4}    & \multicolumn{2}{c}{Z3}       & \multicolumn{2}{c}{MathSAT} \\ \cline{4-9}
%			&  & & \# goals validated & avg.   & \# goals validated & avg.    & \# goals validated & avg.   \\ \hline
%			\multirow{2}{*}{included}& fp.sqrt    & 7 & 7  & 39.123 & 0  & -       & -  & -      \\
%			& axiomatized sqrt & 7 & 5  & 27.197 & 1  & 1.146   & -  & -      \\ \hline
%			\multirow{2}{*}{removed} & fp.sqrt    & 7 & 7  & 45.299 & 4  & 169.015 & 5  & 1.244  \\
%			& axiomatized sqrt & 7 & 5  & 27.251 & 5  & 32.443  & 5  & 19.388 \\ \hline
%		\end{tabular}
%%	\end{adjustbox}
%	\caption{Summary statistics for benchmarks containing the square root function, with and without quantified axioms in the SMT-LIB translation }
%	\label{Tab:sqrtWithWithoutQuantifer}
%\end{table*}

All the solvers we have used in this work comply with the IEEE 754 standard
and therefore have bit-precise support for the square root function.
They provide bit-precise reasoning by effectively encoding the behavior of floating-point circuits over bitvectors (which is naturally expensive), together with different heuristics and abstractions to speed up solving time.
However, depending on the property, we
do not always need bit-precise reasoning, so
we propose handling the square root function with the same taclet-based
axiomatization as introduced in \autoref{sec:taclets-key}.

To this end, we conducted an experiment on the benchmarks containing
\texttt{sqrt}, comparing the approach from \autoref{sec:taclets-key} (adding
the necessary axioms, resp. taclet rules) to using the square root implemented
in SMT solvers (\texttt{fp.sqrt}). We chose to include only axioms specified in or inferred from the IEEE 754 standard (e.g. if the argument of the square root function is NaN or less than zero, then the square root results in NaN). The full set of axioms that we used is included in the Appendix of the technical report.

Rows 7 and 8 in~\autoref{Tab:smtResultsWithWithoutAxioms} summarize the results for this experiment;
the detailed results
are included in the Appendix of the technical report.
We observed that for two out of the three benchmarks,  the average running time of all solvers decreases using the axiomatized square root.
Furthermore, Z3 is able to reason about more
proof obligations with the axiomatized version. However, the success of this approach depends on the axioms
added to \KeY and may not always work if we do not have suitable axioms.  For example, for the \texttt{Circuit.instantCurrent}
benchmark (\autoref{circuit-bench}), using the axiomatized square root, CVC4 is not able
to validate the contract, but with \texttt{fp.sqrt} the contract is validated.

In summary, treating \texttt{sqrt} axiomatically can result in shorter solving
times than performing bit-precise reasoning, but the approach may not always
succeed when the axioms are not sufficient to prove a particular property.

\subsection{Discussion and insights}
The experiments show that highly automated floating point program verification is viable for relevant properties (handling of special values and some functional properties), up to a certain level of complexity (given by the SMT solvers). The choices of which parts of a proof obligation are delegated to SMT, and how they are translated to SMT, are crucial for achieving effective and efficient program verification. Arithmetic operations proved to be more efficiently dealt with by delegation to SMT, whereas for transcendental functions, axiomatization and rule based treatment in the theorem prover, outside the SMT solver, performs clearly better.

% \begin{lstlisting}[float, caption=The Circuit.instantCurrent benchmark, label=spurious-example]
%  /*@ public normal_behaviour
%    @  requires !\fp_nan(d)&& !\fp_infinity(d);
%    @  ensures !\fp_nan(\result);
%    @*/
% public static double method(double d){
%    return Math.sin(d);
% }
% \end{lstlisting}

% !TEX root = main.tex
\section{Related Work}

Our implementation uses the floating-point SMT-LIB theory~\cite{Brain2015},
% which is today supported by several SMT solvers~\cite{Z32008,Brain2019,Brain2014}, but which
which however does not handle
transcendental functions, as their semantics is (library)
implementation dependent. Some real-valued automated solvers do handle
transcendental functions~\cite{dReal2013,Akbarpour2010}, but to the best of our
knowledge, the combination of floating-points and reals in SMT solvers is still
severely limited.

%\subsection{Deductive Verification}
None of the existing deductive verifiers support floating-point
transcendental functions automatically.
The Why3 deductive verification framework~\cite{Why32013}
has support for floating-point arithmetic, with front-ends for
the C and Ada programming languages through Frama-C~\cite{Frama-C-2012} and
SPARK~\cite{ChapmanSchanda2014,Fumex2017}, respectively. Why3 has back-end support for
different SMT solvers, as well as interactive proof assistants like Coq.
Until recently, Why3 would discharge still many interesting floating-point
problems with help of Coq, relying on significant user interaction. In
later work~\cite{Fumex2017} (in the context with floating-point verification
for Ada programs), Why3 can achieve a higher degree of automation. Note,
however, that the user is still required to add code assertions as well as
`ghost code' to a significant extent.

The Boogie intermediate verification language~\cite{leino2008boogie} also
supports floating-point expressions, and targets Z3 for discharging proof
obligations. In the Boogie community, it was observed that writing a
specification in Boogie leads to decreases in SMT solver performance when
compared to writing the goal in SMT-LIB directly, probably due to an inherent
mixing of theories when using
Boogie~\cite{BoogieIssue2019}. %\footnote{\url{https://github.com/boogie-org/boogie/issues/109}}. %
This matches our own experiences, and separation of theories should be
considered an important task for the further development of floating-point
verification.

Other deductive verifiers for Java have only rudimentary support for
floating-points. Verifast~\cite{DBLP:conf/nfm/JacobsSPVPP11} treats
floating-point operations as if they were real values, and OpenJML~\cite{openjml}
parses programs with floating-point operations, but essentially treats
\texttt{float} and
\texttt{double} as uninterpreted sorts.

The Java category of verification competition
SV-COMP~\cite{SVCOMP2020} contains a number of benchmarks that make
use of floating-point variables. However, the focus of these benchmarks
is usually not on arithmetical properties of expressions, but on the
completeness of the Java language support.
%
% \subsection{Model Checking Java Programs}
Amongst the participants of SV-COMP~2020,
the Symbolic (Java) Pathfinder
(SPF)~\cite{DBLP:conf/issta/PasareanuMBGLPP08} (and various extensions)
and the Java Bounded Model
Checker (JBMC)~\cite{DBLP:conf/cav/CordeiroKKST18} support
floating-point arithmetic.  Besides being limited to exploring the
state space up to a bounded depth,
%SPF does not support symbolic
%values (since it is an explicit-state model checker), and JBMC's
their constraint languages do not support quantifiers and abstracting of
method calls---which are features that we have used in this work.

% Java Pathfinder~\cite{DBLP:journals/ase/VisserHBPL03} is an
% explicit-state model checker for the formal analysis of Java
% programs. It employs a special purpose virtual machine that operates
% on concrete values, but whenever a nondeterministic choice is made,
% stores remaining choices in a backlog to process them later. While
% the virtual machine can operate on concrete floating-point values,
% it does not deal with symbolic values and, thus, cannot be used to
% analyze the kind of questions that we deal with in this paper.
%
% The Java Bounded Model Checker
% (JBMC)~\cite{DBLP:conf/cav/CordeiroKKST18} checks runtime exceptions
% and user-defined assertions by unwinding loops and inlining method
% calls (up to a predefined depth) and passing the resulting formula
% to a decision procedure.
% %
% JBMC can deal with floating-point values in programs and assertions
% and many of the examples in our collection could also be formulated
% as JBMC input. However, its constraint language does not support
% quantifiers, and method calls cannot be abstracted in
% contracts---two features that we used in this work.

%\subsection{Interactive Verification and Formalisations}
Floating-point arithmetic has also been formalized in several interactive
theorem provers~\cite{Jacobsen2015,Boldo2011,Fox2017}. While one can prove
intricate properties about floating-point
programs~\cite{Boldo2013,Boldo2009,Harrison2000}, proofs using interactive
provers are to a large part manual and require significant expertise.

%\subsection{Abstract Interpretation and Static Analysis}
Abstract interpretation based techniques can show the absence of special
values in floating-point code fully automatically, and several abstract
domains which are sound with respect to floating-point arithmetic
exist~\cite{Chen2008,Jeannet2009}. While the analysis itself is fully
automated, applying it successfully to real-world programs in general
requires adaptation to each program analyzed by end-users, e.g. the
selection of suitable abstract domains or widening
thresholds~\cite{Blanchet2003}.

Besides showing the absence of special values, recent research has developed
static analyses to bound floating-point roundoff
errors~\cite{Fluctuat2011,Daisy2018,FPTaylor2015,real2Float2017,Precisa2017}.
These analyses currently work only for small arithmetic kernels and the
tools in particular do not accept programs with objects.

%\subsection{Dynamic Analyses}
Dynamic analyses generally scale well on real-world programs, but can only
identify bugs (when given failure-triggering input),
rather than proving correctness for \emph{all} possible inputs.
Executing a floating-point program together
with a higher-precision one allows one to find inputs which cause large
roundoff errors~\cite{Benz2012,Chiang2014,Lam2013}.
Ariadne~\cite{Barr2013} uses a combination of symbolic execution,
real-valued SMT solving and testing to find inputs that trigger
floating-point exceptions, including overflow and invalid operations. Our
work subsumes this approach as the SMT solvers that we use can directly
generate counterexamples, but more importantly, KeY is able to prove the
absence of such exceptions.

%%% Local Variables:
%%% mode: latex
%%% TeX-master: "main"
%%% End:

% !TEX root = main.tex
\section{Conclusion}

By joining the forces of rule-based deduction and SAT-based SMT solving, we
presented the first working floating-point support in a deductive verification
tool for Java and by that close a remaining gap in KeY to now support full
sequential Java.
Our evaluation shows that for specifications dealing with value ranges and
absence of NaN and infinity, our approach can verify realistic programs within a
reasonable time frame.
We observe that the MathSAT and CVC4 solver's floating-point support scales
sufficiently for our benchmarks, as long as the queries do not include any
quantifiers, and that our axiomatized approach for handling transcendental
functions is best realized using calculus rules in \KeY's internal reasoning
engine.
While our work is implemented within the \KeY{} verifier, we expect our approach
to be portable to other verifiers.

% We have collected a set of real-world code challenges with floating-points
% from a variety of sources and have evaluated our approach on
% this benchmark.
% %
% For specifications dealing with value
% ranges and absence of NaN and infinity, our
% approach can verify programs within a
% reasonable time frame.
% (a few minutes)\mattias{even better, I'd  say. How to formulate?}.
% Eva: I think we don't need to be too specific.
%
% We also evaluated the
% running times of the different solvers. MathSAT closes more proof obligations,
% but CVC4 is faster. Quantifiers slow down solvers considerably and reduce
% their ability to find counterexamples.
% % \mattias{@Rosa. Do you agree. Probably the MathSAT bit is not true when I look at the graphs.}\rosa{@Mattias: it is the best in terms of number of goals it can validate and invalidate. But CVC4 is the best in terms of average runtime}
% Our approach also supports programs and specifications that use transcendental functions.
% Adding quantified axioms to the translated SMT code has
% proven to be less satisfactory. Instead we realise the axioms as calculus rules
% for \KeY's reasoning engine, allowing us to close more proof obligations
% successfully.
% By joining the forces of rule-based deduction and SAT-based SMT solving,
% we have closed a remaining gap in a verification system to now support full sequential Java.

% While our work is implemented within the \KeY{} verifier, we expect
% our approach to be portable to other verifiers.

\subsection*{Acknowledgements}
This research was partially funded by the Deutsche Forschungsgemeinschaft (DFG, German Research Foundation) project 387674182.
The authors would like to thank Daniel Eddeland, who together with co-author W.~Ahrendt performed prestudies which impacted the current work.

%%
%% Bibliography
\bibliographystyle{splncs04}{}
\bibliography{biblio_clean}

%\clearpage
% Eva: for Arvix version, comment out the Open Access Text!
\newpage
\appendix
% !TEX root = main.tex
\section{Appendix}\label{Sec:appendix}
\subsection{Axioms for Transcendental Functions in KeY}
Here we present the axioms that we implemented to prove properties for benchmarks with transcendental functions:

\begin{itemize}
	\item	If \texttt{arg} is NaN or an infinity, then \texttt{sin(arg)} is NaN.
	\item	If \texttt{arg}  is zero, then the result of \texttt{sin(arg)}  is a zero with the same sign as \texttt{arg}.
	\item if \texttt{arg} is not NaN or infinity, then the returned value of
	\texttt{sin} is between $-1.0$ and $1.0$.
	\item if \texttt{arg} is not NaN or infinity, then the returned value of
	\texttt{sin} is not NaN.
	\item	If \texttt{arg} is NaN or an infinity, then \texttt{cos(arg)} is NaN.
	\item if \texttt{arg} is not NaN or infinity, then the returned value of
	\texttt{cos} is between $-1.0$ and $1.0$.
	\item if \texttt{arg} is not NaN or infinity, then the returned value of
	\texttt{cos} is not NaN.
	\item	If \texttt{arg} is NaN or an infinity, then \texttt{atan(arg)} is NaN.
	\item	If \texttt{arg}  is zero, then the result of \texttt{atan(arg)}  is a zero with the same sign as \texttt{arg}.
	\item if \texttt{arg} is not NaN, then the returned value of
	\texttt{atan} is between $-\pi/2$ and $\pi/2$.

\end{itemize}

In our Evaluation we showed that handling square root axiomatically can improve performance. Here is the list of axioms we used for this function:

\begin{itemize}
\item	If \texttt{arg} is NaN or less than zero, then \texttt{sqrt(arg)} is NaN.
\item	If \texttt{arg} is positive infinity, then \texttt{sqrt(arg)} is positive infinity.
\item	If \texttt{arg} is positive zero or negative zero, then  \texttt{sqrt(arg)} is the same as \texttt{arg}.
\item	If \texttt{arg}  is not NaN and greater or equal to zero, then  \texttt{sqrt(arg)} is not NaN.
\item	If \texttt{arg} is not infinity and is greater than one then   \texttt{sqrt(arg)} $<$ \texttt{arg}.
\end{itemize}

\subsection{Detailed Evaluation Results}
Here we present the tables that did not fit in the main body of the paper and contain  detailed results of our experiments. In each table we show the number of goals per benchmark that each solver can validate or invalidate, together with the average and maximum time (in seconds) needed. `TO' in the maximum column denotes that at least one goal timed out.
The goals resulting in timeout were excluded from the computation of the average time.

\autoref{Tab:smtResultsWithAxioms} shows the
	results for benchmarks with valid contracts with the quantified formulas  included in the SMT translations.  We have summarized this table in row 1 of \autoref{Tab:smtResultsWithWithoutAxioms}.
\begin{table}[p]
	\centering
	%\begin{adjustbox}{width=1\textwidth}
		\renewcommand{\arraystretch}{1}
\begin{tabular}{lccrrcrr}
	\toprule
	\multirow{2}{*}{benchmark} & \multirow{2}{*}{\# goals} & \multicolumn{3}{c}{CVC4}                                                & \multicolumn{3}{c}{Z3}                                                                     \\ \cline{3-8}
	&                           & \begin{tabular}[c]{@{}c@{}}\# goals\\ proven\end{tabular} & avg. & max. & \begin{tabular}[c]{@{}r@{}}\# goals\\ proven\end{tabular} & avg.                    & max. \\ \midrule
	Complex.add(1)             & 1                         & 1                                                         & 0.6  & 0.6  & 1                                                         & 1.6                     & 1.6  \\
	Complex.divide(1)          & 2                         & 2                                                         & 1.7  & 1.9  & 2                                                         & 1.8                     & 2.3  \\
	Complex.divide(2)          & 8                         & 8                                                         & 3.4  & 11.5 & 4                                                         & 3.2                     & TO   \\
	Complex.compare            & 2                         & 2                                                         & 0.5  & 0.6  & 2                                                         & 1.5                     & 1.5  \\
	Complex.reciprocal(1)      & 2                         & 2                                                         & 1.0  & 1.7  & 0                                                         & -                       & TO   \\
	Complex.reciprocal(2)      & 2                         & 2                                                         & 1.8  & 2.4  & 2                                                         & 2.7                     & 4.0  \\
	Circuit.impedance          & 4                         & 4                                                         & 0.8  & 0.9  & 3                                                         & 87.5                    & TO   \\
	Circuit.current(1)         & 4                         & 4                                                         & 6.3  & 13.2 & 0                                                         & -                       & TO   \\
	Circuit.current(2)         & 1                         & 1                                                         & 5.5  & 5.5  & 0                                                         & -                       & TO   \\
	Matrix2.transposedEq       & 1                         & 1                                                         & 0.5  & 0.5  & 0                                                         & -                       & TO   \\
	Matrix3.transposedEqV2     & 1                         & 1                                                         & 1.3  & 1.3  & 0                                                         & -                       & TO   \\
	Rectangle.scale(1)         & 32                        & 31                                                        & 2.2  & TO   & 7                                                         & 46.6                    & TO   \\
	Rotate.computeError        & 8                         & 8                                                         & 9.8  & 13.9 & 0                                                         & -                       & TO   \\
	Rotate.computeRelErr       & 8                         & 8                                                         & 25.3 & 45.6 & 0                                                         & -                       & TO   \\
	FPLoop.fploop              & 4                         & 4                                                         & 0.5  & 1.0  & 4                                                         & \multicolumn{1}{r}{2.3} & 8.1 \\ \bottomrule
\end{tabular}
	%\end{adjustbox}
	\caption{Summary of valid goals proved and running times of each solver for the SMT translations \emph{with} quantified axioms}
	\label{Tab:smtResultsWithAxioms}
\end{table}

\autoref{Tab:smtResultsWithoutAxioms} demonstrates the results for the same benchmarks when the quantified axioms are removed form the SMT translations which is summarized in row 2 of   \autoref{Tab:smtResultsWithWithoutAxioms}.
\begin{table}[p]
	\centering
	\begin{adjustbox}{width=1\textwidth}
		\renewcommand{\arraystretch}{1.2}
\begin{tabular}{lccrrcrrcrr}
	\toprule
	\multirow{2}{*}{benchmark} & \multirow{2}{*}{\# goals} & \multicolumn{3}{c}{CVC4}                                                & \multicolumn{3}{c}{Z3}                                                   & \multicolumn{3}{c}{MathSAT}                                              \\ \cline{3-11}
	&                           & \begin{tabular}[c]{@{}c@{}}\# goals\\ proven\end{tabular} & avg. & max. & \begin{tabular}[c]{@{}c@{}}\# goals\\ proven\end{tabular} & avg. & max.  & \begin{tabular}[c]{@{}c@{}}\# goals\\ proven\end{tabular} & avg. & max.  \\ \midrule
	Complex.add(1)             & 1                         & 1                                                         & 0.5  & 0.5  & 1                                                         & 1.3  & 1.3   & 1                                                         & 0.2  & 0.2   \\
	Complex.divide(1)          & 2                         & 2                                                         & 1.6  & 1.9  & 2                                                         & 2.1  & 2.7   & 2                                                         & 1.5  & 1.9   \\
	Complex.divide(2)          & 8                         & 8                                                         & 3.4  & 11.4 & 4                                                         & 2.5  & TO    & 8                                                         & 29.0 & 209.5 \\
	Complex.compare            & 2                         & 2                                                         & 0.5  & 0.6  & 2                                                         & 1.1  & 1.3   & 2                                                         & 0.2  & 0.2   \\
	Complex.reciprocal(1)      & 2                         & 2                                                         & 0.9  & 1.5  & 1                                                         & 198.5 & TO    & 2                                                         & 2.2  & 2.4   \\
	Complex.reciprocal(2)      & 2                         & 2                                                         & 1.7  & 2.3  & 2                                                         & 2.8  & 3.4   & 2                                                         & 1.5  & 1.9   \\
	Circuit.impedance          & 4                         & 4                                                         & 0.7  & 0.7  & 4                                                         & 9.5  & 17.1  & 4                                                         & 0.4  & 0.5   \\
	Circuit.current(1)         & 4                         & 4                                                         & 6.3  & 13.0 & 0                                                         & -    & TO    & 4                                                         & 13.3 & 36.2  \\
	Circuit.current(2)         & 1                         & 1                                                         & 5.2  & 5.2  & 0                                                         & -    & TO    & 1                                                         & 26.3 & 26.3  \\
	Matrix2.transposedEq       & 1                         & 1                                                         & 0.5  & 0.5  & 0                                                         & -    & TO    & 1                                                         & 0.6  & 0.6   \\
	Matrix3.transposedEqV2     & 1                         & 1                                                         & 1.1  & 1.1  & 0                                                         & -    & TO    & 1                                                         & 3.9  & 3.9   \\
	Rectangle.scale(1)         & 32                        & 31                                                        & 2.1  & TO   & 32                                                        & 95.1 & 251.8 & 32                                                        & 2.4  & 4.2   \\
	Rotate.computeError        & 8                         & 8                                                         & 9.7  & 13.5 & 0                                                         & -    & TO    & 8                                                         & 22.7 & 35.7  \\
	Rotate.computeRelErr       & 8                         & 8                                                         & 25.0 & 45.0 & 0                                                         & -   & TO    & 8                                                         & 27.5 & 46.4  \\
	FPLoop.fploop              & 4                         & 4                                                         & 0.5  & 1.1  & 4                                                         & 2.0  & 7.2   & 4                                                         & 0.4  & 1.0   \\ \bottomrule
\end{tabular}
	\end{adjustbox}
	%\captionsetup{width=1\textwidth}
	\caption{Summary of valid goals proved and running times of each solver for the SMT translations \emph{without} quantified axioms}
	\label{Tab:smtResultsWithoutAxioms}
\end{table}

\autoref{Tab:smtResultsCEWithWithoutAxiomsDetailed} shows the detailed results of the experiments with benchmarks with invalid contracts, when the quantified formulas are included in and removed form the SMT translations.  This results are summarized in rows 3 and 4 of \autoref{Tab:smtResultsWithWithoutAxioms}.
\begin{table}[t]
	\centering
	\begin{adjustbox}{width=1\textwidth}
		\renewcommand{\arraystretch}{1.2}
\begin{tabular}{lccccrrccrrccrr}
	\toprule
	\multirow{3}{*}{benchmark}         & \multicolumn{2}{c}{\multirow{2}{*}{\# goals}} & \multicolumn{4}{c}{CVC4}                                                     & \multicolumn{4}{c}{Z3}                                                       & \multicolumn{4}{c}{MathSAT}                                                  \\ \cline{4-15}
	& \multicolumn{2}{c}{}         & \multicolumn{2}{c}{\# goals} & \multirow{2}{*}{avg.} & \multirow{2}{*}{max.} & \multicolumn{2}{c}{\# goals} & \multirow{2}{*}{avg.} & \multirow{2}{*}{max.} & \multicolumn{2}{c}{\# goals} & \multirow{2}{*}{avg.} & \multirow{2}{*}{max.} \\
	& valid                        & invalid        & valid        & invalid       &                       &                       & valid        & invalid       &                       &                       & valid        & invalid       &                       &                       \\ \midrule
	\textit{with quantified axioms}    &                              &                &              &               &                       &                       &              &               &                       &                       &              &               &                       &                       \\
	Matrix3.transposedEq               & 0                            & 1              & 0            & 0             & -                     & TO                    & 0            & 0             & -                     & TO                    & -            & -             & -                     & -                     \\
	Rectangle.scale(2)                 & 12                           & 4              & 12           & 0            & 12.2                  & TO                    & 8            & 0             & 4.6                   & TO                    & -            & -             & -                     & -                     \\
	Complex.add(2)                     & 2                            & 2              & 2            & 0             & 0.6                   & 0.7                   & 2            & 0             & 1.4                   & TO                    & -            & -             & -                     & -                     \\
	FPLoop.fploop2                     & 3                            & 1              & 3            & 0             & 0.9                   & 1.7                   & 3            & 0             & 0.5                   & TO                    & -            & -             & -                     & -                     \\
	FPLoop.fploop3                     & 3                            & 1              & 3            & 0             & 0.4                   & 1.7                   & 3            & 0             & 0.3                   & TO                    & -            & -             & -                     & -                     \\ \midrule
	\textit{without quantified axioms} &                              &                &              &               &                       &                       &              &               &                       &                       &              &               &                       &                       \\
	Matrix3.transposedEq               & 0                            & 1              & 0            & 1             & 170.2                 & 170.2                 & 0            & 0             & -                     & TO                    & 0            & 1             & 16.2                  & 16.2                  \\
	Rectangle.scale(2)                 & 12                           & 4              & 12           & 3             & 12.2                  & TO                    & 12           & 3             & 108.2                 & TO                    & 12           & 4             & 2.4                   & 9.5                   \\
	Complex.add(2)                     & 2                            & 2              & 2            & 2             & 0.5                   & 0.5                   & 2            & 2             & 0.7                   & 1.0                   & 2            & 2             & 0.2                   & 0.2                   \\
	FPLoop.fploop2                     & 3                            & 1              & 3            & 1             & 0.4                   & 0.6                   & 3            & 1             & 0.9                   & 1.7                   & 3            & 1             & 0.3                   & 0.5                   \\
	FPLoop.fploop3                     & 3                            & 1              & 3            & 1             & 0.3                   & 0.6                   & 3            & 1             & 0.6                   & 1.7                   & 3            & 1             & 0.2                   & 0.4                   \\ \bottomrule
\end{tabular}
	\end{adjustbox}
	\caption{Summary of invalid goals proved and running times of each solver for the SMT translations with and without quantified axioms}
	\label{Tab:smtResultsCEWithWithoutAxiomsDetailed}
\end{table}
%\item
%\autoref{Tab:axoimResultsDetailed} compares the results from applying the two
%approaches for handling transcendental functions in sections \ref{Sec:axiomInSolver} and \ref{sec:taclets-key}, using the default SMT translation in \KeY. This table is summarized in rows 5 and 7 of  \autoref{Tab:smtResultsWithWithoutAxioms}.
%
%\item
%\autoref{Tab:axiomInKeyWithoutQuantifierDetailed}, depicts the results of applying the approach in \autoref{sec:taclets-key},  while the quantified formulas are removed from the SMT translations. This table is summarized in row 8 of \autoref{Tab:smtResultsWithWithoutAxioms}.

\begin{table}[t]
	\centering
	\begin{adjustbox}{width=1\textwidth}
		\renewcommand{\arraystretch}{1.1}
		\begin{tabular}{lccrrcrrcrr}
			\toprule
			\multirow{2}{*}{benchmark}                & \multirow{2}{*}{\# goals} & \multicolumn{3}{c}{CVC4}                                                         & \multicolumn{3}{c}{Z3}                                                           & \multicolumn{3}{c}{MathSAT}                                                      \\ \cline{3-11}
			&                           & \begin{tabular}[c]{@{}c@{}}\# goals\\ validated\end{tabular} & avg.    & max.    & \begin{tabular}[c]{@{}c@{}}\# goals\\ validated\end{tabular} & avg.    & max.    & \begin{tabular}[c]{@{}c@{}}\# goals\\ validated\end{tabular} & avg.    & max.    \\ \hline
			\emph{fp.sqrt}                                     &                           &                                                              &         &         &                                                              &         &         &                                                              &         &         \\
			Cartesian.toPolar                 & 4                         & 4                                                            & 6.9   & 7.5   & 1                                                            & 23.5 & TO      & 4                                                            & 1.2   & 1.7   \\
			Cartesian.distanceTo              & 1                         & 1                                                            & 8.2   & 8.2   & 0                                                            & - & TO & 1                                                            & 1.0   & 1.0   \\
			Circuit.instantCurrent & 2                         & 2                                                            & 123.5 & 127.5 & 0                                                            & -       & TO      & 0                                                            & -       & TO      \\ \midrule
			\emph{axiomatized sqrt}                            &                           &                                                              &         &         &                                                              &         &         &                                                              &         &         \\
			Cartesian.toPolar                 & 4                         & 4                                                            & 2.0   & 2.9   & 4                                                            & 49.81  & 163.0  & 4                                                            & 1.0   & 1.6   \\
			Cartesian.distanceTo              & 1                         & 1                                                            & 2.7  & 2.7   & 1                                                            & 233.0  & 233.0  & 1                                                            & 1.0  & 1.0   \\
			Circuit.instantCurrent & 2                         & 0                                                     & -  & TO & 0                                                            & -       & TO      & 0 (2 CE)                                                     & 11.1 & 13.8 \\ \bottomrule
		\end{tabular}
	\end{adjustbox}
	\caption{Summary statistics for benchmarks containing the square root function, with quantified formulas removed from the SMT-LIB translation }
	\label{Tab:sqrtWithoutQuantifer}
\end{table}

The first two sections of \autoref{Tab:axiomInKeyWithWithoutQuantifierDetailed} show the results from applying the two
approaches for handling transcendental functions in sections \ref{Sec:axiomInSolver} and \ref{sec:taclets-key}, using the default SMT translation in \KeY. The last section of the table depicts the results of applying the approach in \autoref{sec:taclets-key},  while the quantified formulas are removed from the SMT translations.
This table is summarized in rows 5 and 6 of  \autoref{Tab:smtResultsWithWithoutAxioms}
\begin{table}[t]
	\centering
	\begin{adjustbox}{width=1\textwidth}
		\renewcommand{\arraystretch}{1.1}
\begin{tabular}{lccrrcrrcrr}
	\toprule
	\multirow{2}{*}{benchmark} & \multirow{2}{*}{\# goals} & \multicolumn{3}{c}{CVC4}                                                                                   & \multicolumn{3}{c}{Z3}                                                                                     & \multicolumn{3}{c}{MathSAT}                                                                                        \\ \cline{3-11}
	&                           & \begin{tabular}[c]{@{}c@{}}\# goals\\ validated\end{tabular} & avg.                 & max.                 & \begin{tabular}[c]{@{}c@{}}\# goals\\ validated\end{tabular} & avg.                 & max.                 & \begin{tabular}[c]{@{}l@{}}\# goals\\ validated\end{tabular} & \multicolumn{1}{r}{avg.} & \multicolumn{1}{r}{max.} \\ \midrule
	\multicolumn{11}{l}{\emph{axioms in SMT-LIB translation}}                                                                                                                                                                                                                                                                                                                                            \\
	Cartesian.toPolar          & 4                         & 4                                                            & 7.1                  & 9.2                 & 1                                                            & 16.8                    & TO                   & -                                                            & -                        & -                        \\
	Polar.toCartesian          & 2                         & 2                                                            & 0.9                  & 0.9                  & 2                                                            & 69.7                    & 95.7                   & -                                                            & -                        & -                        \\
	Circuit.instantCurrent     & 2                         & 1                                                            & 123.6                & TO                   & 0                                                            & -                    & TO                   & -                                                            & -                        & -                        \\
	Circuit.instantVoltage     & 2                         & 2                                                            & 1.1                  & 1.1                  & 1                                                            & 103.8                    & TO                   & -                                                            & -                        & -                        \\ \midrule
	\multicolumn{11}{l}{\emph{axioms as taclet rules in \KeY with quantified formulas}}                                                                                                                                                                                                                                                                                                   \\
	Cartesian.toPolar          & 4                         & 4                                                            & 6.8                  & 7.7                  & 1                                                            & 40.0                    & TO                   & -                                                            & -                        & -                        \\
	Polar.toCartesian          & 2                         & 2                                                            & 1.4                  & 1.9                  & 1                                                            & 288.0                    & TO                   & -                                                            & -                        & -                        \\
	Circuit.instantCurrent     & 2                         & 2                                                            & 123.8                & 128.3                & 0                                                            & -                    & TO                   & -                                                            & -                        & -                        \\
	Circuit.instantVoltage     & 2                         & 2                                                            & 1.3                  & 1.3                  & 0                                                            & -                    & TO                   & -                                                            & -                        & -                        \\ \midrule
	\multicolumn{11}{l}{\emph{axioms as taclet rules in \KeY without quantified formulas}}                                                                                                                                                                                                                                                                                                \\
	Cartesian.toPolar           & 4                         & 4                                                            & 6.9   & 7.5   & 1                                                            & 23.5 & TO     & 4                                                            & 1.2 & 1.7 \\
	Polar.toCartesian          & 2                         & 2                                                            & 1.5   & 2.3   & 2                                                            & 52.6   & 81.2 & 2                                                            & 0.6 & 0.8 \\
	Circuit.instantCurrent    & 2                         & 2                                                            & 123.5 & 127.5 & 0                                                            & -       & TO     & 0                                                            & -     & TO    \\
	Circuit.instantVoltage      & 2                         & 2                                                            & 1.5   & 1.7   & 2                                                            & 146.4 & 160.7 & 2                                                            & 0.8 & 0.8    \\ \bottomrule
\end{tabular}
	\end{adjustbox}
	\caption{Summary statistics with axioms in SMT-LIB translations and  as taclet rules in \KeY}
	\label{Tab:axiomInKeyWithWithoutQuantifierDetailed}
\end{table}

\autoref{Tab:sqrtWithoutQuantifer} shows the detailed results of conducting the experiment on the benchmarks containing
\texttt{sqrt}, comparing the approach from \autoref{sec:taclets-key} (adding
the necessary axioms, resp. taclet rules) to using the square root implemented in SMT solvers (\texttt{fp.sqrt}), when the quantified formulas are removed from the SMT translations.
We have summarized the results of these experiments in rows 7 and 8 of \autoref{Tab:smtResultsWithWithoutAxioms}.

\subsubsection{Sensitivity to Contract Variations}\label{Sec:variations}

We conducted an experiment on our \texttt{Rectangle.scale} benchmark
to assess the solver's sensitivity to various changes,
applied to the benchmark's contract or its implementation. We considered
the following modifications:
\begin{itemize}
	\item $v0$: is the original version of the benchmark (\autoref{rectangle-bench} using the second contract) and our baseline;
	\item $v1$: reduces the number of classes involved to two, while keeping the same functionality;
	\item $v2$: reduces the number of classes involved to one, while keeping the same functionality;
	\item $v3$: modifies $v2$ such that variable bounds in the precondition become more ``complicated''
	in terms of longer fractional parts (e.g. the bounds for \texttt{arg2} become
	[3.0000001, -6.4000000003] instead of [3.0001, -6.4000003]);
	\item $v4$: simplifies the mathematical expression of $v2$ (less arithmetic operations)
	\item $v5$: modifies $v3$ such that \texttt{arg2} has a tighter bound, i.e. the interval width is smaller
	\item $v6$: modifies $v2$ such that \texttt{arg2} has a larger bound, i.e. the interval width is larger
	\item $v7$: modifies $v2$ such that only \texttt{arg2} has a ``complicated'' bound
	\item $v8$: modifies $v0$ such that \texttt{arg2} has a tighter bound
\end{itemize}

\autoref{Tab:rectangleWithAxiomDetailed}
% and \ref{Tab:rectangleWithoutAxiomDetailed}
summarizes the results for this experiment.
With the quantified forulas included in the SMT translation, Both CVC4 and Z3 are able to prove more goals when the number of classes is
reduced, and also when the number of arithmetic operations is reduced. Z3
further seems to be sensitive to whether variable bounds are ``complicated''
or not, whereas CVC4 is not.
We obtain a somewhat surprising result when \texttt{arg2} has a tighter
bound. While Z3's performance improves, CVC4 validates two goals less. On the
other hand, increasing the bounds on \texttt{arg2} does not seem to make a
difference.

It seems that \texttt{arg2} is the bottleneck for this benchmark; when only
\texttt{arg2} has a ``complicated'' input interval, CVC4 proves less
goals. Finally, constraining \texttt{arg2} in the original benchmark more
tightly allows CVC4 to validate all goals but Z3's performance remains unaffected.

%TODO we dont have the table for this anymore should we keep it?
With the quantified formulas
removed from the SMT-LIB translation, we can see that CVC4's results, in
terms of number of goals validated, are the same as before, and Z3 performs
much better than before. MathSAT is able to validate all goals of all versions.

In summary, solvers' performance seems to be sensitive to slight innocuous
looking changes such as the number of classes involved and variable bounds.
This behavior could be potentially exploited by e.g. relaxing a variable's
bounds.

\begin{table*}[t]
	\centering
	\begin{adjustbox}{width=\textwidth}
		\renewcommand{\arraystretch}{1.1}
		\begin{tabular}{llccrrcrr}
			\thickhline
			\multirow{2}{*}{version} & \multirow{2}{*}{applied change}          & \multirow{2}{*}{\# goals} & \multicolumn{3}{c}{CVC4}                                                     & \multicolumn{3}{c}{Z3}                                                        \\ \cline{4-9}
			&                                          &                           & \begin{tabular}[c]{@{}l@{}}\# goals\\ validated\end{tabular} & avg.  & max.  & \begin{tabular}[c]{@{}l@{}}\# goals\\ validated\end{tabular} & avg.    & max. \\ \hline
			v0                       & none (original)                          & 32                        & 31                                                           & 2.3 & TO    & 7                                                           & 46.2  & TO   \\
			v1                       & fewer classes (2) in v0                  & 32                        & 32                                                           & 2.5 & 4.8    & 9                                                           & 42.7  & TO   \\
			v2                       & fewer classes (1) in v0                  & 32                        & 32                                                           & 2.4 & 4.2    & 7                                                           & 85.2 & TO   \\
			v3                       & complicated intervals for all vars in v2 & 32                        & 32                                                           & 2.5 & 5.5 & 6                                                           & 59.3  & TO   \\
			v4                       & simpler math in v2                       & 16                        & 16                                                           & 1.0 & 1.3 & 12                                                           & 35.3  & TO   \\
			v5                       & shorter interval for arg2 in v3          & 32                        & 30                                                           & 2.3 & TO    & 9                                                           & 95.1  & TO   \\
			v6                       & longer interval for arg2 in v2           & 32                        & 32                                                           & 2.6 & 7.4 & 7                                                           & 14.8  & TO   \\
			v7                       & complicated interval for arg2 in v2      & 32                        & 31                                                           & 2.4 & TO    & 7                                                           & 23.9  & TO   \\
			v8                       & shorter interval for arg2 in v0          & 32                        & 32                                                           & 2.5 & 4.2 & 7                                                           & 46.5 & TO   \\ \thickhline
		\end{tabular}
	\end{adjustbox}
	\caption{SMT solvers summary statistics for various versions of the \texttt{ Rectangle}  benchmark with quantified axioms in the SMT translations}
	\label{Tab:rectangleWithAxiomDetailed}
\end{table*}

\end{document}